%% file: SDHCALDigitizer.tex
\pdfoutput=1
\documentclass{JINST}

\usepackage{cite}
\usepackage{lineno}
\usepackage{url}
\usepackage{enumerate}
\usepackage{graphicx}
\usepackage{caption}
\usepackage{subcaption}
\usepackage{amsmath}
\usepackage[frenchb,english]{babel}

\title{Resistive Plate Chamber Digitization in a Hadronic Shower Environment}

\input{AuthorList}

\abstract{The CALICE Semi-Digital Hadron Calorimeter technological prototype is a sampling calorimeter using Glass Resistive Plate Chamber detectors with a three-threshold readout as the active medium. This technology is one of the two options proposed for the hadron calorimeter of the International Large Detector for the International Linear Collider. The prototype was exposed to beams of muons, electrons and pions of different energies at the CERN Super Proton Synchrotron. To be able to study the performance of such a calorimeter in future experiments it is important to ensure reliable simulation of its response. In this paper we present our prototype simulation performed with GEANT4 and the digitization procedure achieved with an algorithm called SimDigital. A detailed description of this algorithm is given and the methods to determinate its parameters using muon tracks and electromagnetic showers are explained. The comparison with hadronic shower data shows a good agreement up to 50~GeV. Discrepancies are observed at higher energies. The reasons for these differences are investigated.}

\begin{document}
%\linenumbers

%%%%%%%%%%%%%%%%%%%%%%%%%%%%%%%%%%%%%%%%%%%%%%%%%%%%%%%%%%%%%%%%%%%%%%%%%%%%%%%%%%%%%%%%%%%%%%%%%%

\section{Introduction}
\label{sec:intro}
The CALICE Semi-Digital Hadron Calorimeter~(SDHCAL) technological prototype was built in 2011 \cite{main}. It was designed to provide a powerful tool for hadronic energy measurement and for the application of the Particle Flow Algorithm for the detectors of the future International Linear Collider (ILC). The SDHCAL is a high granularity sampling calorimeter with 48 Glass Resistive Plate Chambers~(GRPC) used as active media with a transversal size of 1~m$^2$ divided into $96\times96$ readout cells of 1~cm$^2$ each. Absorber layers are made of 2~cm thick stainless steel plates. The longitudinal size is about 1.3~m which leads to a total depth of about 6$\lambda_I$ for the SDHCAL prototype. 

It has been shown that hadronic calorimeter prototypes using GRPCs as an active material provide a precise energy measurement over a wide energy range with either binary or semi-digital readout \cite{jose,energyReco}. 
The SDHCAL prototype is also a useful tool to track particles in hadronic showers by identifying segments using tracking techniques such as Hough Transform as it has been shown in \cite{sdhcalCalor}.
Moreover, the GEANT4 Collaboration has been developing models to simulate hadronic showers for years \cite{geant4}. These models have been evaluated by different experiments \cite{atlas,ahcal-geant} in which transversal segmentation was not as fine as the one of the SDHCAL prototype. This calorimeter may thus help to constrain these models. However, the simulation of Resistive Plate Chambers (RPC) response to hadronic showers is not trivial. Unlike the case with muon detectors, where resistive plate chambers are commonly used, many charged particles from showers can cross the gas gap simultaneously leading, in some cases, to saturation effects for which the RPC's response needs to be correctly modeled.

This paper presents a digitization method to transform the simulated energy deposited by the passage of charged particles through the gas, into a semi-digital information. The simulated response is compared to that obtained using the SDHCAL prototype. It is structured as follows: in section \ref{sec:grpc}, a brief description of the GRPC used for the SDHCAL is given. Section \ref{sec:method} explains the different steps of the SDHCAL simulation and digitization whereas section \ref{sec:params} presents the method used to determine the parameters introduced in section \ref{sec:method}. Finally, section \ref{sec:results} shows comparisons between data and a few hadronic shower models used in GEANT4.

%%%%%%%%%%%%%%%%%%%%%%%%%%%%%%%%%%%%%%%%%%%%%%%%%%%%%%%%%%%%%%%%%%%%%%%%%%%%%%%%%%%%%%%%%%%%%%%%%% 

\section{Description of Glass Resistive Plate Chambers}
\label{sec:grpc}
%\graphicspath{{figs/grpc/}}
The active detectors of the SDHCAL are 1~m$^2$ Glass Resistive Plate Chambers. The cathode and the anode are glass plates with thicknesses of 1.1~mm and 0.7~mm respectively. These electrodes are painted with a resistive coating on the outer surfaces which allows to apply high voltage ($\simeq$ 7 kV). The gas gap between the two electrodes is 1.2~mm. The readout layer is divided in $96\times96$ pick-up pads of $10\times10$~mm$^2$, separated by 0.406~mm. The gas mixture is 93$\%$ of TetraFluoroEthane (TFE), 5$\%$ of CO$_2$ and 2$\%$ of SF$_6$ \cite{main}. The TFE is the main gas and was chosen for its low ionisation level. The CO$_2$ and the SF$_6$ are quenchers: they limit the size of the charge avalanche, and they reduce the rate of avalanches due to thermal and other sources of noise \cite{grpc-gas}.

When a charged particle crosses the gas gap, several gas molecules are ionized\footnote{The average number of primary ionisations is around 10 along the gas gap for particles crossing the chamber perpendicularly.}. Ions and electrons are then accelerated by the strong electric field created by the high voltage applied on the electrodes. The electrons ionize other gas molecules. An avalanche is then created. The signal on the pads is recorded by HARDROC2 ASICs \cite{duluc} in a 2-bit format, corresponding to three thresholds related to the amount of induced charge. These three thresholds were initially set at 0.114, 5.0 and 15.0~pC. The aim of these thresholds is to obtain additional information on the number of particles crossing the pad and to improve the hadronic shower energy measurement, as described in \cite{energyReco}. Several pads can be fired when only a charged particle crosses the gas gap. This so-called pad multiplicity will be an important element to be discussed in sections \ref{sec:method} and \ref{sec:params}.

\begin{figure}[t]
  \begin{center}
    \includegraphics[width=0.7\textwidth]{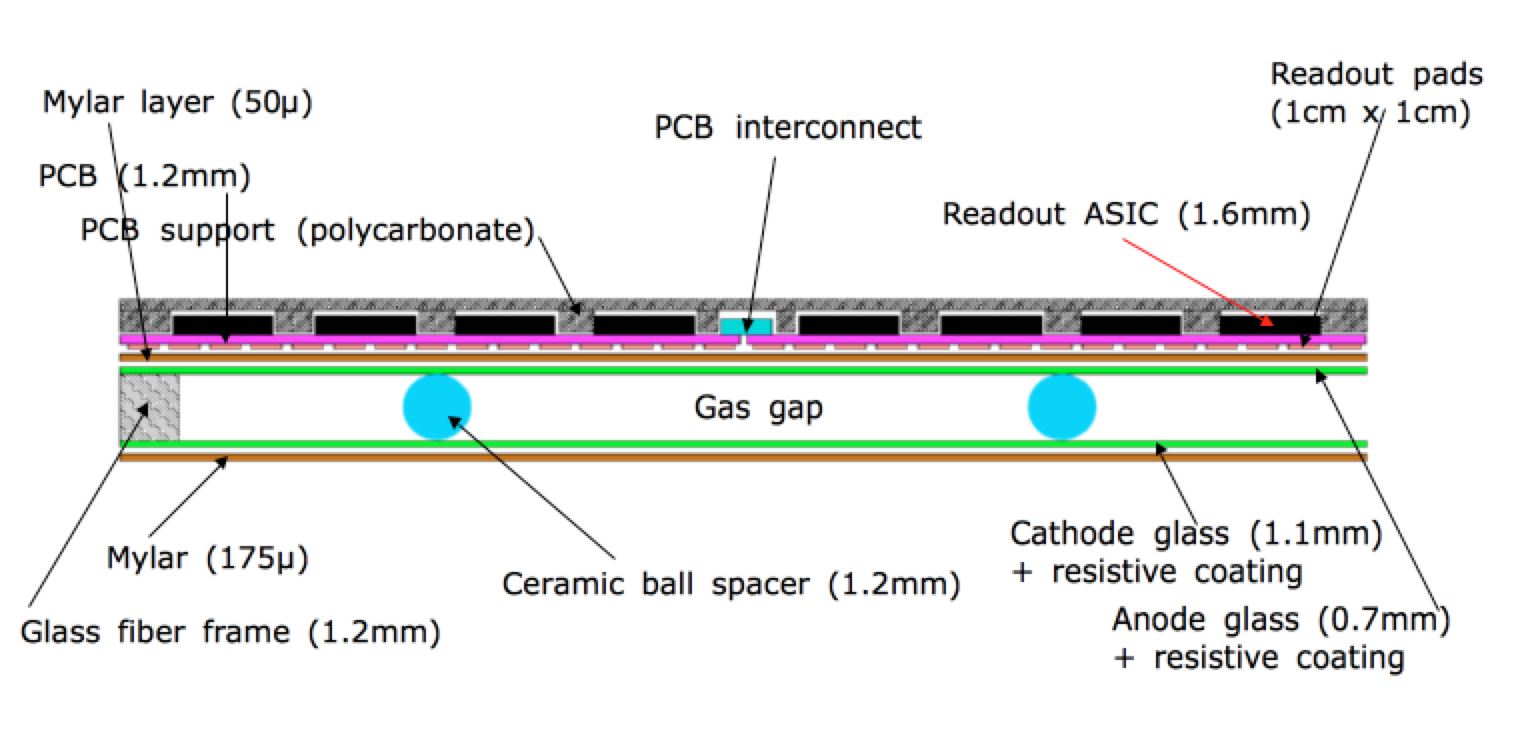}
    \caption{A schematic cross-section of a SDHCAL active layer (not to scale).}
    \label{fig.grpcK7}
  \end{center}
\end{figure}
A schematic cross section of one glass resistive plate chambers is shown in Fig.~\ref{fig.grpcK7}.
In the SDHCAL prototype, GRPCs are operated in avalanche mode. This mode is described in \cite{abbrecia} where it is shown that a Polya distribution could be used to simulate the amount of charge, $q$, deposited in the anode. The Polya distribution is given by the following equation:
\begin{equation}
  \label{eq.polya}
  P(q)=\frac{1}{\Gamma(1+\delta)}\left(\frac{1+\delta}{\bar q}\right)^{1+\delta}q^\delta e^{[-\frac{q}{\bar q}(1+\delta)]}
\end{equation}
where $\bar q$ is the average value of the deposited charge in the anode, $\delta$ is related to the width of the distribution and $\Gamma$ is the Gamma function. 

%%%%%%%%%%%%%%%%%%%%%%%%%%%%%%%%%%%%%%%%%%%%%%%%%%%%%%%%%%%%%%%%%%%%%%%%%%%%%%%%%%%%%%%%%%%%%%%%%% 

\section{SDHCAL simulation and digitization method}
\label{sec:method}
%\graphicspath{{figs/method/}}
The SDHCAL prototype simulation is performed with a program based on the GEANT4 toolkit~\cite{geant4} where each SDHCAL element is described using its composition, density, size and position. Pion, electron, proton and muon events are simulated using different physics lists prepared by the GEANT4 collaboration. A physics list defines the different GEANT4 models and their transitions used to simulate physical processes. In this paper, the QGSP\_BERT\_HP and FTFP\_BERT\_HP physics lists are used to simulate hadronic and electromagnetic showers in the SDHCAL prototype using the 9.6 GEANT4 version. These two physics lists are those recommended by the GEANT4 collaboration.
In addition to the GEANT4 based program, a new algorithm called SimDigital is developed to perform the digitization. In GEANT4, the energy deposited in the gas is recorded whereas in data the induced charge is measured. The multiplicity effect is also not included in GEANT4. The SDHCAL simulation output contains the following information: the list of steps\footnote{A step in GEANT4 is a segment of a particle path. In addition, each time the particle meets a material boundary or has an interaction a new step is created.} inside the gas gaps; the deposited energy in these steps; the start and end point positions of each step in gas gaps; and the occurrence time of each step in the gap. It may happen that GEANT4 produces several steps for only one particle inside the gas gap. To avoid the simulation of several avalanches for only one particle in the gap, these steps are linked together before writing the simulation output. However, one may expect that for particles with large angle with respect to the normal to GRPC's, triggering multiple avalanches in the gas gap should be allowed. This will be take into account by correcting the induced charge using the steps angle (see Eq.~\ref{eq.lengthcorrection} and section~\ref{sec.angle-correction}).

The SimDigital algorithm is implemented as a Marlin \cite{marlin} processor in the  MarlinReco \cite{marlinreco} package of ILCSoft \cite{ilcsoft}. The aim of the SimDigital algorithm is to determine the induced charge from each particle crossing a gas gap, to distribute this charge over the pick-up pads and to apply the thresholds. It is formulated according to the following sequence: 
\begin{enumerate}[~~1-]
\item During beam tests, no external trigger system was used. The hits from showers incident particles, muons, cosmics and noise were recorded using a 200~ns clock and an event building procedure was needed. For each time slot that contains more than seven hits, hits belonging to neighbouring time slots are added to those of the central one to build one physics event. More details are given in \cite{main}. In this study, five time slots are used to aggregate physics event (1000 ns). Thus, a signal from late interacting particles like neutrons might not be included in the event. To take this into account, all steps recorded after 1000~ns from the primary particle time generation are rejected. 
\item \label{it.1st} One pad ($P_0$) where one or several charged particles are crossing the gas gap is selected. The length of each step generated inside the gas gap is calculated. For example, if the particle trajectory is perpendicular to the GRPCs, the step maximum length corresponds to the gap distance (1.2~mm).
\item The length of some steps inside the gas gap could be almost zero. This can randomly happen during the particle propagation by GEANT4. However, this occurs quite often in the vicinity of the detector's boundaries. Figure~\ref{fig.map_and_length_vs_deltaz}(a) shows the step length versus the difference ($\Delta_z$) between the middle position of the step and the middle of the gas gap. This figure shows that a large fraction of zero length steps is located near the gas gap boundary ($|\Delta_z|\simeq$ 0.6~mm). To avoid charge avalanches from these non physical zero length steps, those with a length smaller than a given value $l_{min}$ are rejected.
  \begin{figure}[t]
    \centering
    \begin{subfigure}[b]{.49\linewidth}
      \includegraphics[width=1.0\textwidth]{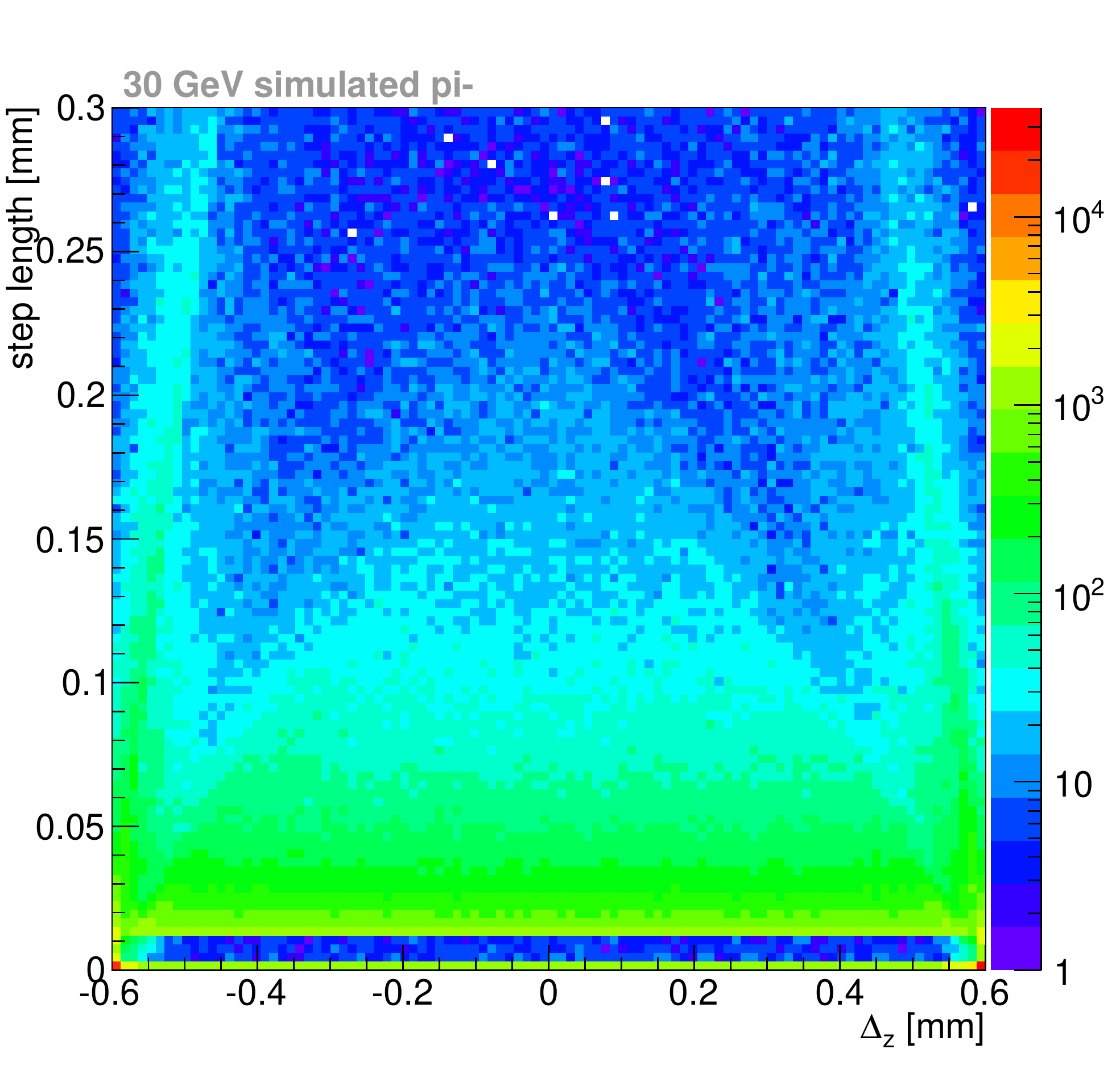}
      \caption{ }
    \end{subfigure}
    \begin{subfigure}[b]{.49\linewidth}
      \includegraphics[width=1.0\textwidth]{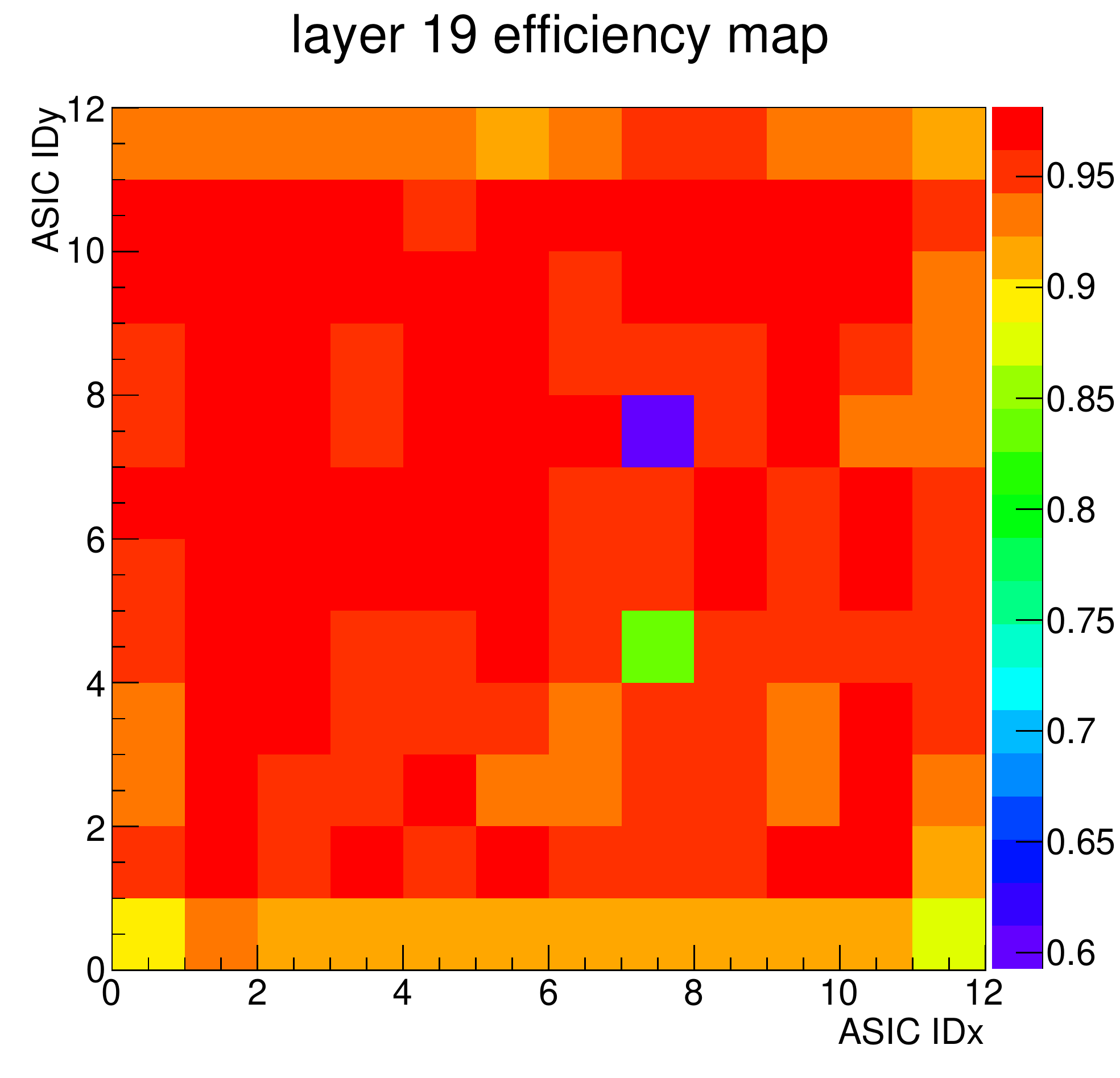}
      \caption{ }
    \end{subfigure}
    \caption{(a): Length of steps in~mm as a function of $\Delta_z$ in~mm. This figure is zoomed on the short steps region to show that most of the short steps are located at the detector's boundaries ($|\Delta_z|\simeq$ 0.6~mm). (b): Measured SDHCAL layer ASIC efficiency map example. Each bin corresponds to the mean efficiency measured in one ASIC which corresponds to $8\times8$ readout pads.}
    \label{fig.map_and_length_vs_deltaz}
  \end{figure}
\item The prototype measured efficiency maps\footnote{Efficiency per ASIC is estimated from muon data with the method described in \cite{energyReco}.} are used to select the steps. Figure~\ref{fig.map_and_length_vs_deltaz}(b) shows an example of one layer efficiency map. If steps are located in a region for which the prototype efficiency is $90\%$, $10\%$ of them are dropped randomly. This allows us to take into account the effect of quenchers not included in GEANT4 and to avoid having simulated hits in dead or masked electronic channels.
\item An induced charge ($q$) is randomly chosen for each selected step using the Polya distribution defined by Eq.~\ref{eq.polya}. This induced charge is then corrected as follows:
  \begin{equation}
    \label{eq.lengthcorrection}
    q_{Corrected} = \left\{ \begin{array}{rl}
      q\left(\frac{d_s}{d_{gap}}\right)^\kappa &\mbox{ if $\frac{d_s}{d_{gap}}>1$} \\
      q &\mbox{ otherwise}
    \end{array} \right.
  \end{equation}
  where $d_s$ is the step length, $d_{gap}$ the size of the gap (1.2~mm) and $\kappa$ is a free parameter. When the step is crossing the whole gap, the fraction $\frac{d_s}{d_{gap}}$ is equivalent to $\frac{1}{\cos\theta}$, where $\theta$ is the angle between the normal to the GRPC's plan and the step. The effect of such a correction will be discussed in the next section.
\item When two ionizing particles are close, their induced avalanches may overlap but the detected signal is not equivalent to the sum of the two avalanches. So if two steps are closer than a given distance $d_{cut}$ the step with the lowest induced charge is rejected\footnote{More realistic simulations of the charge screen effect could be envisaged but would require more parameters to tune.}.
\item \label{chargespitting} The charge ratios between $P_0$ and its neighbouring pads is then estimated to account for the multiplicity effect. The neighbouring pads are the pads in the same layer at a distance (from pad center to pad center) smaller than a given distance $r_{max}$ from $P_0$. Those charge ratios $R_i$ are defined through a sum of Gaussian functions:
  \begin{equation}
    \label{eq.ratio}
    R_i = \frac{\int_{a_i}^{b_i}\int_{c_i}^{d_i}\sum_{j=0}^{n}\alpha_j e^{ \frac{(x_0-x)^2+(y_0-y)^2}{2\sigma_j^2}}dxdy}{N}
  \end{equation}
  where $a_i,\ b_i,\ c_i,\ d_i$ represent the border positions of the pad $i$ that are within $r_{max}$ from $P_0$. $x_0$ and  $y_0$ are the step centre coordinates and $N$ is the normalisation factor defined as: 
  \begin{equation}
    \label{eq.norm}
    N=\int_{-r_{max}}^{+r_{max}}\int_{-r_{max}}^{+r_{max}}\sum_{j=0}^{n}\alpha_j e^{ \frac{(x_0-x)^2+(y_0-y)^2}{2\sigma_j^2}}dxdy
  \end{equation}
  In Eq.~\ref{eq.ratio} the integer $n$, and the parameters $\alpha_j$, and $\sigma_j$ are free parameters tuned using muon data.
\item The charge of each pad $P_0$ and its neighbours is increased by a factor $R_iq_{Corrected}$.
\item The operation is repeated starting from step \ref{it.1st} for all pads containing at least one step. The collected charge is summed in each pad.
\item Finally, the thresholds are applied for all pads.
\end{enumerate}

To summarise, the SimDigital algorithm introduces several parameters. The Polya distribution parameters ($\bar q$ and $\delta$ in Eq.~\ref{eq.polya}) are determined with a threshold scan on the signal induced by muon tracks. The charge spreading parameters, introduced in Eq.~\ref{eq.ratio} and the charge correction one ($\kappa$ in Eq.~\ref{eq.lengthcorrection}) are estimated to reproduce the pad multiplicity behaviour for single muon tracks. The threshold values are tuned with the efficiency related to each threshold. Finally the parameter $d_{cut}$, used to model the charge screening effect, is tuned to reproduce the number of hits in electromagnetic showers. 

The next section describes the methods used to obtain the best parametrization with the SimDigital algorithm.

%%%%%%%%%%%%%%%%%%%%%%%%%%%%%%%%%%%%%%%%%%%%%%%%%%%%%%%%%%%%%%%%%%%%%%%%%%%%%%%%%%%%%%%%%%%%%%%%%%

\section{Digitizer parameters determination}
\label{sec:params}
%\graphicspath{{figs/params/}}
%%%%%%%%%%%%%%%%%%%%%%%%%%%%%%%%%%%%%%%%%%%%%%%%%
\subsection{Polya distribution}
\label{subsec.thrScan}
\begin{figure}[t]
  \centering
  \begin{subfigure}[b]{.49\linewidth}
    \includegraphics[width=1.0\textwidth]{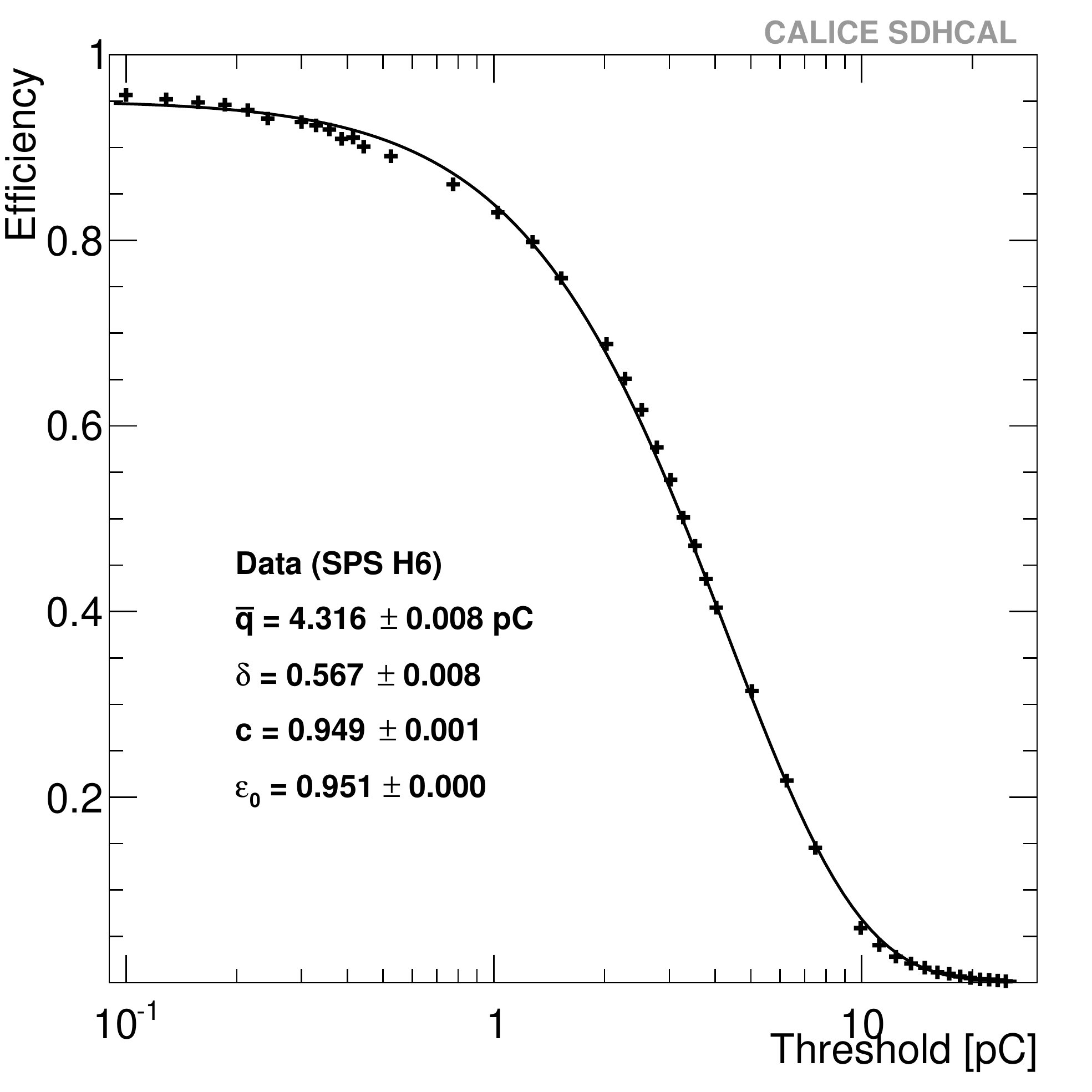}
    \caption{ }
  \end{subfigure}
  \begin{subfigure}[b]{.49\linewidth}
    \includegraphics[width=1.0\textwidth]{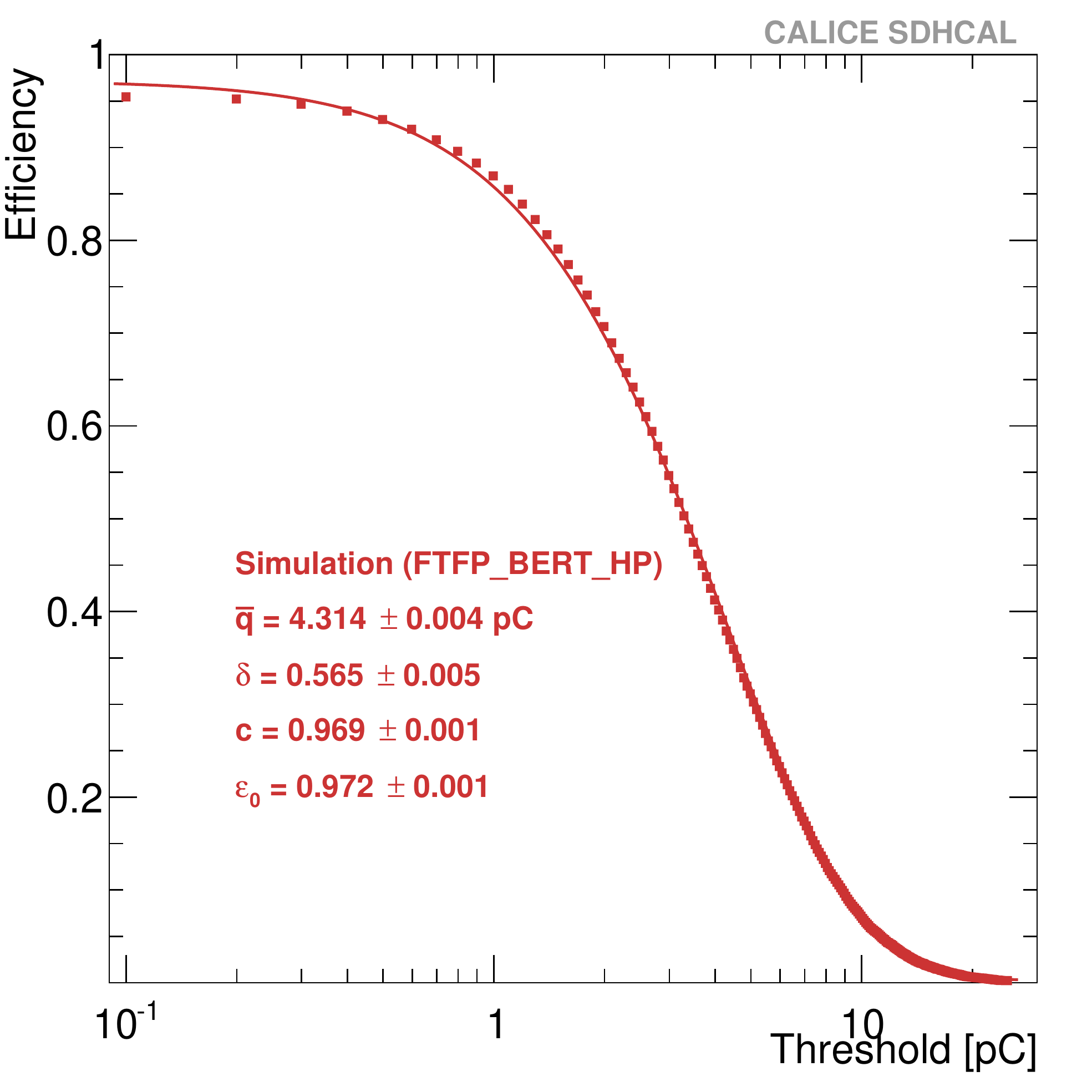}
    \caption{ }
  \end{subfigure}
  \caption{Threshold scan results: average efficiency as a function of threshold for data~(a) and for simulation~(b).}
  \label{fig.thrScan}
\end{figure}

To obtain the Polya parameters (Eq.~\ref{eq.polya}), muon tracks were used to perform a threshold scan study. For this purpose, nine chambers were selected for a dedicated run of the prototype with a muon beam. Different thresholds were applied to the ASICs of these nine layers in order to cover all the induced charge range. The efficiency is computed in the nine chambers as a function of the threshold. To estimate the efficiency in the studied ones, tracks are reconstructed using other chambers located on both sides of the studied chambers.
To build those tracks, hits from one layer are grouped into clusters using a nearest neighbour clustering algorithm\footnote{The distance from one pad center to another one must be smaller than the pitch separating two consecutive pads to gather them into a cluster.}. The clusters' positions are defined with an unweighed barycentre, calculated with the hits' positions. Then a straight trajectory fit is applied (using the clusters' positions) and used to estimate the positions where the track crosses the studied chambers.
A layer is considered efficient if at least one hit is found in a 2.5~cm radius around the expected track impact
\footnote{Only tracks with $\chi^2$ value less than 2 per touched layer are used for this study. The cluster x (resp. y) position uncertainty used to compute the $\chi^2$ is taken as $N_c/\sqrt{12}$ where $N_c$ is the number of the cluster's hits projected on the x (resp. y) axis. Axss x (resp. y) is the horizontal (resp. vertical) axis parallel to the prototype layers.}.
Figure~\ref{fig.thrScan}(a) shows the average efficiency obtained as a function of the threshold.
This curve is then fitted with the integrated Polya function: 
\begin{equation}
  \label{eq.fitScan}
  \varepsilon(q)=\varepsilon_0 - c\int_0^q{\frac{1}{\Gamma(1+\delta)}\left(\frac{1+\delta}{\bar q}\right)^{1+\delta}q'^\delta e^{[-\frac{q'}{\bar q}(1+\delta)]}dq'}
\end{equation}
where $\varepsilon_0$ is the asymptotic value of the efficiency and $c$ is a free parameter. This allows to extract the mean value of the Polya distribution and the width parameter (respectively $\bar q$ and $\delta$ in Eq.~\ref{eq.polya}). The same exercise is performed with the simulation. The Polya parameters $\bar{q}$ and $\delta$ are tuned to reproduce the data efficiency as a function of the thresholds. Figure~\ref{fig.thrScan}(b) presents the simulation threshold scan result. Fit results are shown in Table~\ref{tab.thrScan} for both data and simulation. The value of input Polya parameters used to obtain this result are given in the same table. The digitizer input parameters and the fitted ones after the threshold scan procedure are different. This difference could be explained by the fact that the fit outputs are obtained after eliminating pads whose induced charge is lower than the first threshold value (0.114~pC) and thus inaccessible in this readout scheme.

\begin{table}[t]
  \begin{center}
    \begin{tabular}{c|c|c|c}
      Parameter & Data & Simulation & Digitizer input\\
      \hline
      $\bar q$ & $4.316\pm0.008\,pC$ & $4.314\pm0.004\,pC$ & $4.580\,pC$\\
      $\delta$ & $0.567\pm0.008$ & $0.567\pm0.005$ & $1.120$\\
    \end{tabular}
  \end{center}
  \caption{Measured Polya distribution parameters obtained with a threshold scan.}
  \label{tab.thrScan}
\end{table}
%%%%%%%%%%%%%%%%%%%%%%%%%%%%%%%%%%%%%%%%%%%%%%%%%
\subsection{Charge splitting}
The parameters introduced in Eq.~\ref{eq.ratio} and Eq.~\ref{eq.norm} ($r_{max}$, $n$, $\alpha_j$ and $\sigma_j$) are very important for the charge splitting procedure (step\,\ref{chargespitting} in the SimDigital algorithm). They are tuned to reproduce the muon tracks and the electromagnetic showers responses. The multiplicity which is estimated from the muon tracks response, is defined as the mean number of fired pads in clusters produced by one particle crossing the gas gap. Its average value is estimated using the tracking method described in the previous section. Many different configurations of parameters have been tested to obtain the best parameterization for Eq.~\ref{eq.ratio}. The parameter $n$ (number of Gaussian functions in Eq.~\ref{eq.ratio} and \ref{eq.norm}) was set to 2. It was not possible to reproduce both multiplicity and number of hits in electromagnetic showers using $n=1$. Setting $n=3$ was not found to improve the results. In our optimisation procedure, $r_{max}$ was set to 30~mm\footnote{Beyond this value, the charge contribution is negligible.}. After fixing these parameters, the remaining parameters $\alpha_j$ and $\sigma_j$ were then optimized. Their values are given in table~\ref{tab.summary}.
\begin{figure}[t]
  \begin{center}
    \includegraphics[width=0.49\textwidth]{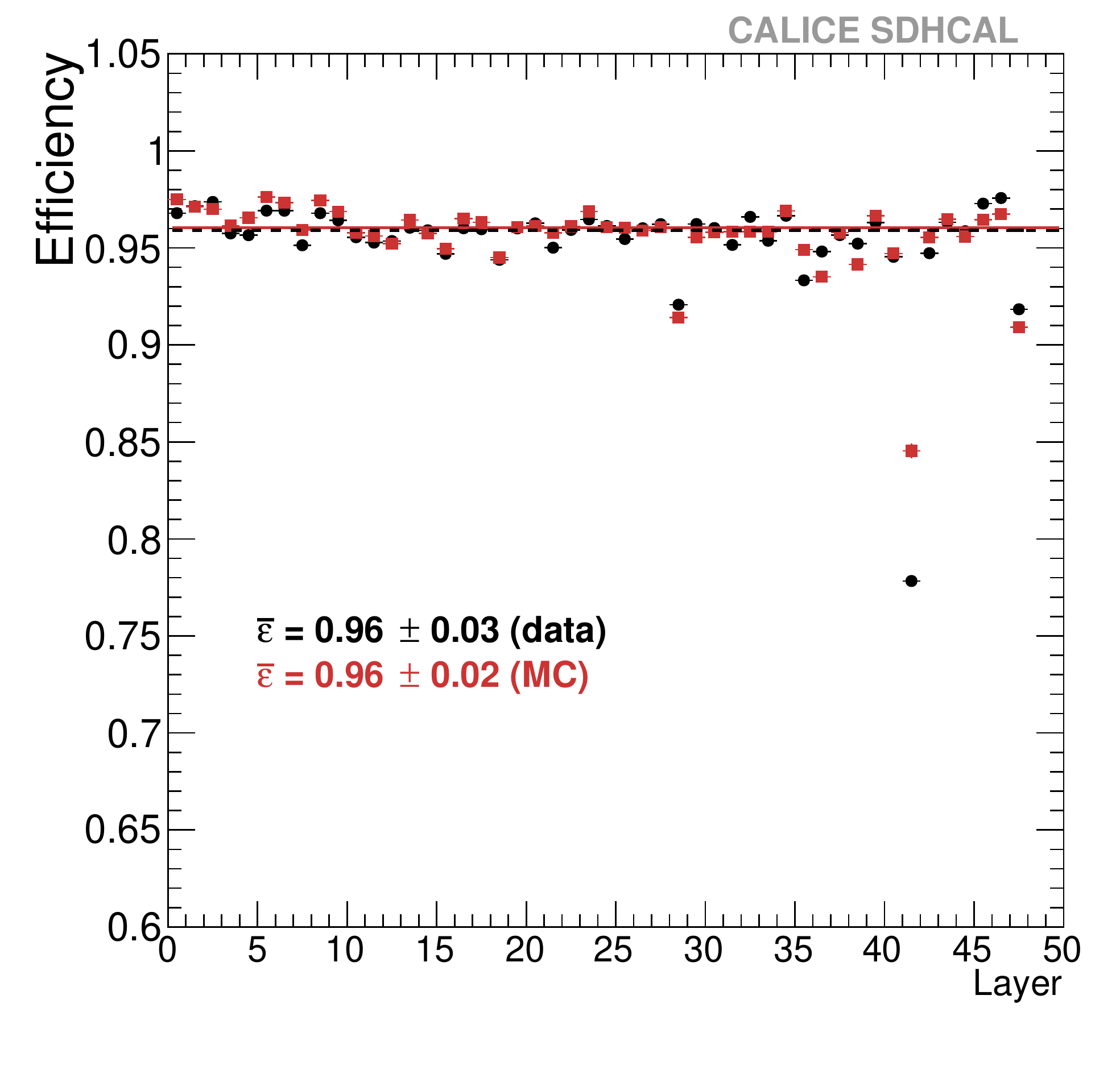}
    \includegraphics[width=0.49\textwidth]{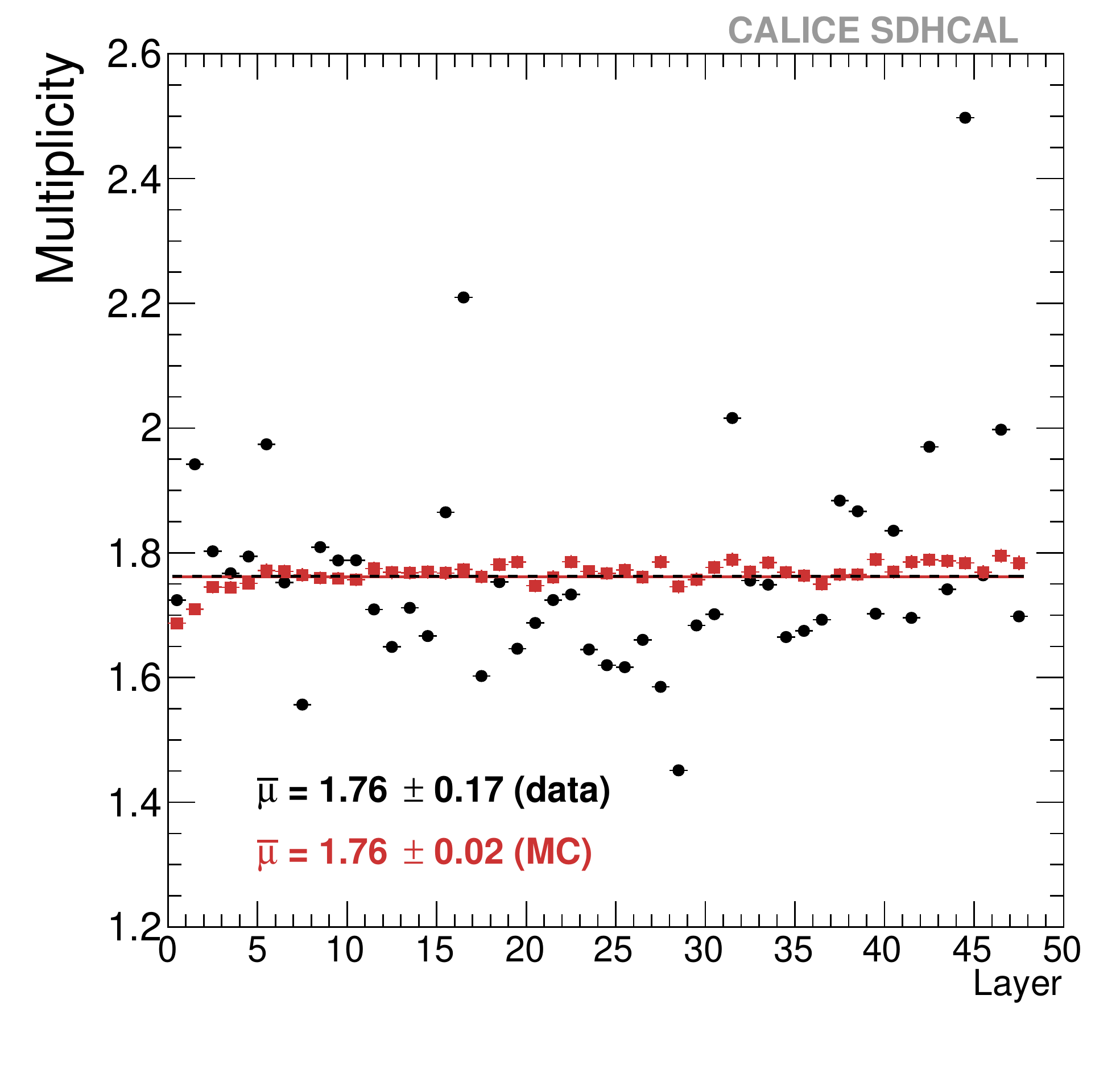}
    \caption{Efficiency (left) and multiplicity (right) per layer with black circles and red squares for beam muon ($\cos\theta\geq0.9$) data and simulation respectively. The lines indicate the average values.}
    \label{fig.eff-mul-layer}
  \end{center}
\end{figure}
Figure~\ref{fig.eff-mul-layer} shows the efficiency and multiplicity per layer for beam muon ($\cos\theta\geq0.9$) data and simulation. The simulated efficiency is closely following the one obtained from data because the efficiency map is included and used in the digitizer. The value of simulated multiplicity is in a good agreement with the data average value. The differences of pad multiplicity from layer to layer in data is most probably due to some differences in the coating resistivity painted on glasses \cite{grpc-gas} and to some imperfections in the gas gap of a few layers which are not taken into account in the simulation.
%%%%%%%%%%%%%%%%%%%%%%%%%%%%%%%%%%%%%%%%%%%%%%%%%
\subsection{Step length correction}
\label{sec.angle-correction}
During the beam tests, the incoming trajectories of muons are perpendicularly incident to the surface of the detectors while in showers, secondary particles can be emitted at various angles. To access the pad multiplicity behavior for particles that are not perpendicular, cosmic muons are used.
\begin{figure}[t]
  \centering
  \begin{subfigure}[b]{.49\linewidth}
    \includegraphics[width=1.0\textwidth]{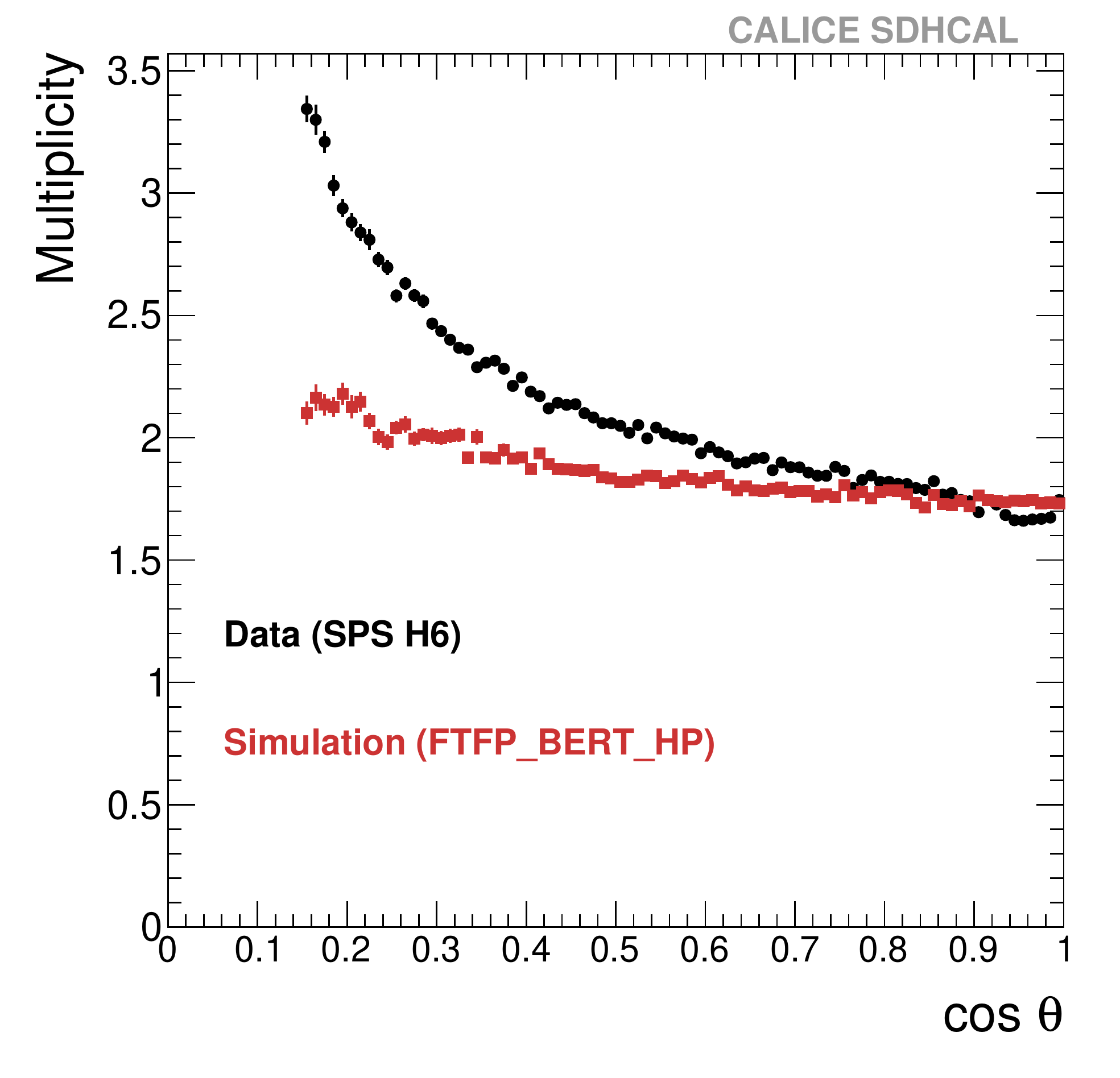}
    \caption{ }
  \end{subfigure}
  \begin{subfigure}[b]{.49\linewidth}
    \includegraphics[width=1.0\textwidth]{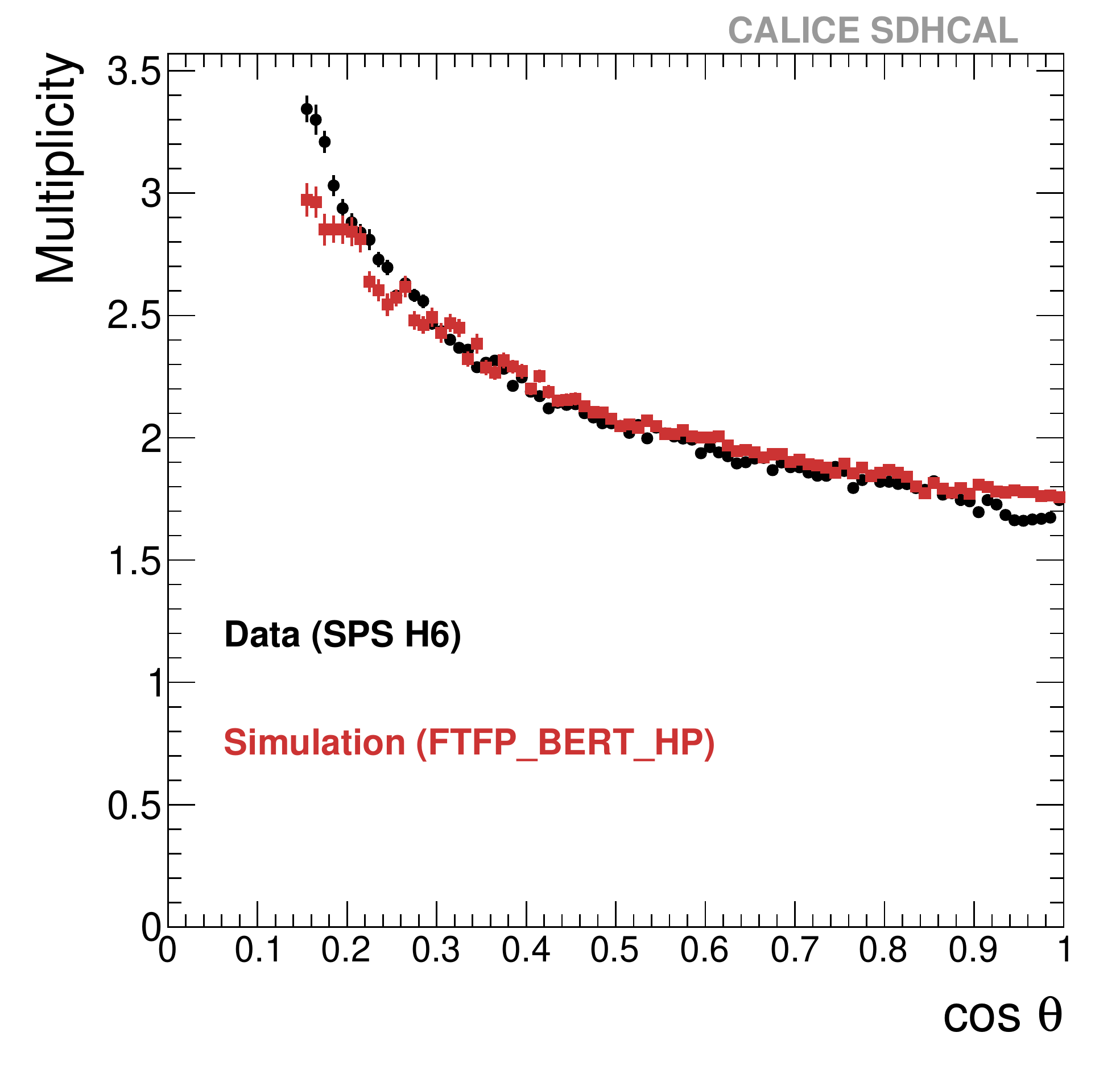}
    \caption{ }
  \end{subfigure}
  \caption{Average pad multiplicity as a function of $\cos\theta$ with black circles and red squares for data and simulation respectively. (a): without digitizer length correction; (b): with digitizer length correction.}
  \label{fig.mul-theta}
\end{figure}
Figure~\ref{fig.mul-theta}(a) shows the pad multiplicity as a function of $\cos\theta$ where $\theta$ is the angle between the normal to the chambers and the reconstructed particle direction. One can see that the multiplicity obtained with data increases with the angle $\theta$ while for the simulation it is flatter\footnote{It is not perfectly flat because the probability of having several steps in the gas increases with increasing angle since in this case the crossed distance in the  gas gap is larger.}. This indicates that an angle correction for the induced charge is needed. A correction depending on $\frac{1}{\cos\theta'}$, where $\theta'$ is the angle between the step and the normal to the chambers, was tested but this introduces singularities when a step is parallel to the detector (in the $(x-y)$ plane).
Therefore Eq.~\ref{eq.lengthcorrection} from section~\ref{sec:method} is used to correct the pad multiplicity with the angle. The best value for the parameter $\kappa$ was found to be 0.4. Figure~\ref{fig.mul-theta}(b) shows a good agreement between data and simulation after applying this correction.
%%%%%%%%%%%%%%%%%%%%%%%%%%%%%%%%%%%%%%%%%%%%%%%%%
\subsection{Threshold tuning}
The three thresholds of the electronic readout are set using a 10-bit Digital Analog Convertor (DAC) for each threshold. Conversion factors between DAC and threshold values are needed for the simulation. To estimate these conversion factors a scan of charge injection was performed on individual ASICs with a dedicated test board\cite{main,these_robert}.
The scan consists in injecting a given charge in the channels of an ASIC and to change the threshold value by steps of 1 DAC unit. The corresponding DAC value ($D_{50\%}$) for which the trigger efficiency is 50$\%$ in each of the channels is then determined. This procedure is then repeated for different injection charge values, for each threshold of the different ASICs. The curve representing the value of $D_{50\%}$ for threshold $i$ as a function of the injected charge is then fitted with a straight line of slope $\lambda_i$.
Finally, to obtain the conversion factor between the DAC value and the charge threshold value, the following equation is used:
\begin{equation}
  T_i = \frac{DAC_i-p_i}{\lambda_i} [pC]
  \label{eq.dacConversion}
\end{equation}
where $T_i$ is the value (in~pC) of the threshold $i$ and $p_i$ is the corresponding average pedestal value. The method to extract the average pedestal value for each threshold is described in \cite{these_robert}. The average values of $\lambda_i$ and the average pedestal values $p_i$ for each threshold are given in Table~\ref{tab.lambdas}.
\begin{table}[!ht]
  \begin{center}
    \begin{tabular}{c|c|c}
      Threshold & $\lambda~[pC^{-1}]$ & Pedestal\\
      \hline
      1 & $700\pm50$ & $90\pm4.5$\\
      2 & $80\pm10$ & $98\pm4.5$\\
      3 & $16.3\pm2$ & $98\pm4.5$
    \end{tabular}
  \end{center}  
  \caption{Average measured conversion factors for each threshold.}
  \label{tab.lambdas}
\end{table}
The values obtained using a board test may differ slightly from those that would have been obtained with the same scan performed on the ASICs embedded on the detector but this was not possible to achieve. Indeed the design of the final printed board circuit does not allow to inject charge. This suggests that slightly different thresholds may be applied in the simulation for a better reproducibility of the observed data. Since the efficiency variation in terms of the lowest threshold was found to be small in the range (threshold$\,\in[0.1,0.4]$~pC as shown in Fig.~\ref{fig.thrScan}(b)), the value of the first threshold in simulation was taken by replacing in Eq.~\ref{eq.dacConversion}, the DAC value used in beam tests ($DAC_0=170$). To fix the second and the third thresholds, the efficiency for those two thresholds is studied. The layer is considered as efficient for threshold 2 (3) if the cluster associated to this layer includes at least one hit exceeding threshold 2 (3). These two threshold values are then tuned to reproduce the related efficiency obtained with muon data.
\begin{figure}[t]
  \centering
  \begin{subfigure}[b]{.49\linewidth}
    \includegraphics[width=1.0\textwidth]{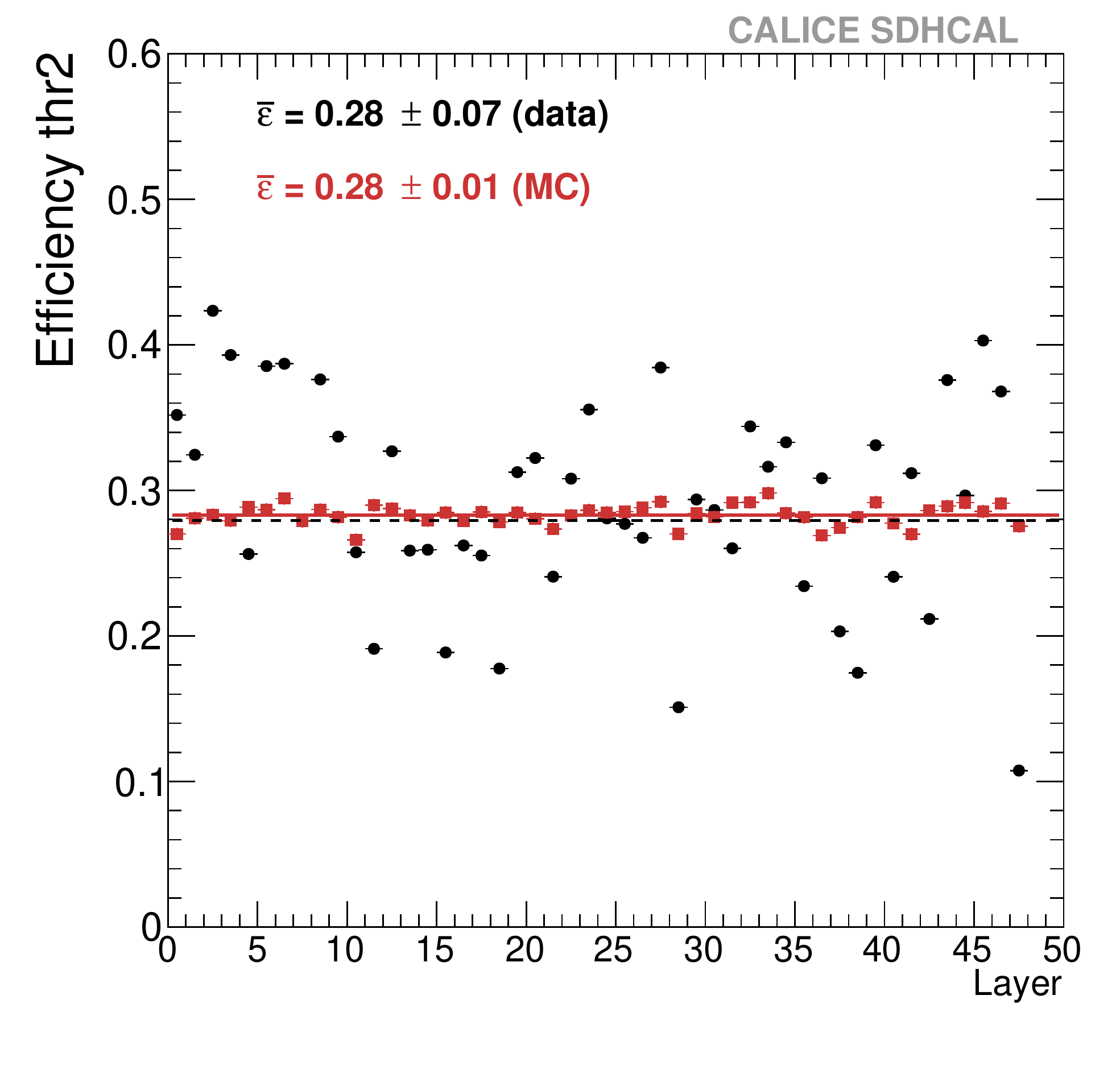}
    \caption{ }
  \end{subfigure}
  \begin{subfigure}[b]{.49\linewidth}
    \includegraphics[width=1.0\textwidth]{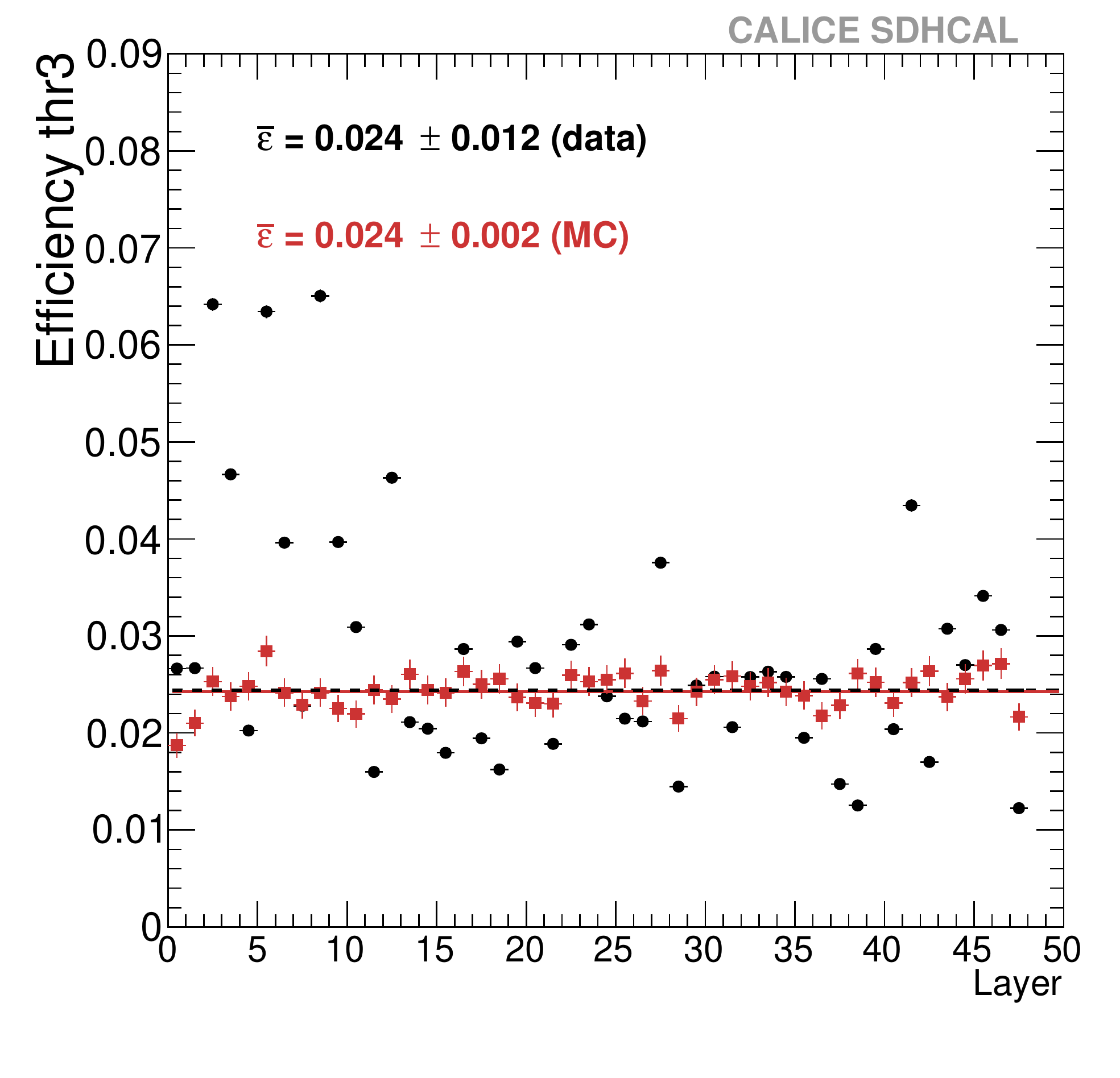}
    \caption{ }
  \end{subfigure}
  \caption{Efficiency per layer for threshold 2 (a) and 3 (b) with black circles and red squares for beam muon ($\cos\theta\geq0.9$) data and simulation respectively.}
  \label{fig.eff_thr}
\end{figure}
Figure~\ref{fig.eff_thr} shows the efficiency per layer for threshold 2 (a) and 3 (b) for both data and simulation. The second threshold is set to 5.4~pC in simulation compared to 5.0~pC in data. The third threshold is set to 14.5~pC in simulation compared to 15.0~pC in data. 
%%%%%%%%%%%%%%%%%%%%%%%%%%%%%%%%%%%%%%%%%%%%%%%%%
\subsection{Other parameters}
\label{sec.parsummary}
\begin{table}[t]
  \begin{center}
    \begin{tabular}{r||l}
      Parameter name & Value \\
      \hline
      \hline
      $l_{min}$ & 0.001~mm\\
      $d_{cut}$ & 0.5~mm \\
      \hline
      $\bar q$ & 4.58~pC \\
      $\delta$ & 1.12 \\ 
      \hline
      $n$ & $2$ \\ 
      $r_{max}$ & 30~mm \\
      $\alpha_0$ & 1.0 \\
      $\sigma_0$ & 1.0~mm \\
      $\alpha_1$ & 0.00083 \\
      $\sigma_1$ & 9.7~mm \\
      \hline
      $\kappa$ & 0.4 \\
      \hline 
      $T_1$ & 0.114~pC\\
      $T_2$ & 5.4~pC \\
      $T_3$ & 14.5~pC
    \end{tabular}
  \end{center}  
  \caption{Digitizer input parameters which allow to obtain the best agreement with data.}
  \label{tab.summary}
\end{table}
The two remaining parameters to be fixed are $l_{min}$ and $d_{cut}$. The parameter $l_{min}$ used to remove zero length steps is set to 1$\mu$m. Variations of this parameter between $0.1$ and 2$\mu$m have negligible effects on the final results of the digitization procedure. The parameter $d_{cut}$ is set to 0.5mm. It is tuned to reproduce the number of hits for electromagnetic showers (see section~\ref{subsec.elec}). \\ \\
Table~\ref{tab.summary} contains digitizer parameters list and their input values.
%%%%%%%%%%%%%%%%%%%%%%%%%%%%%%%%%%%%%%%%%%%%%%%%%%%%%%%%%%%%%%%%%%%%%%%%%%%%%%%%%%%%%%%%%%%%%%%%%%

\section{Digitizer results}
\label{sec:results}
\begin{figure}[t]
  \centering
  \begin{subfigure}[b]{.49\linewidth}
    \includegraphics[width=1.0\textwidth]{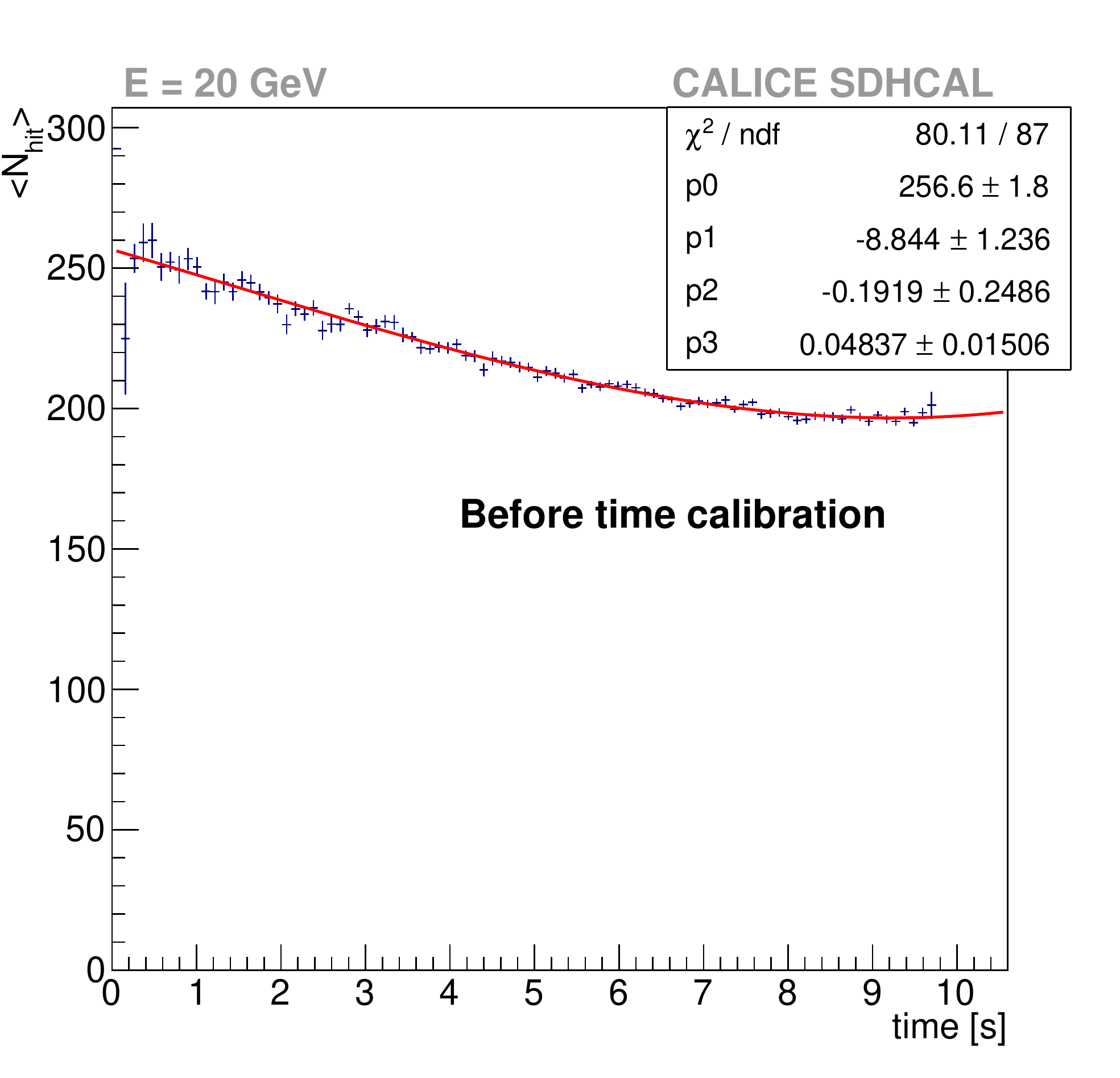}
    \caption{ }
  \end{subfigure}
  \begin{subfigure}[b]{.49\linewidth}
    \includegraphics[width=1.0\textwidth]{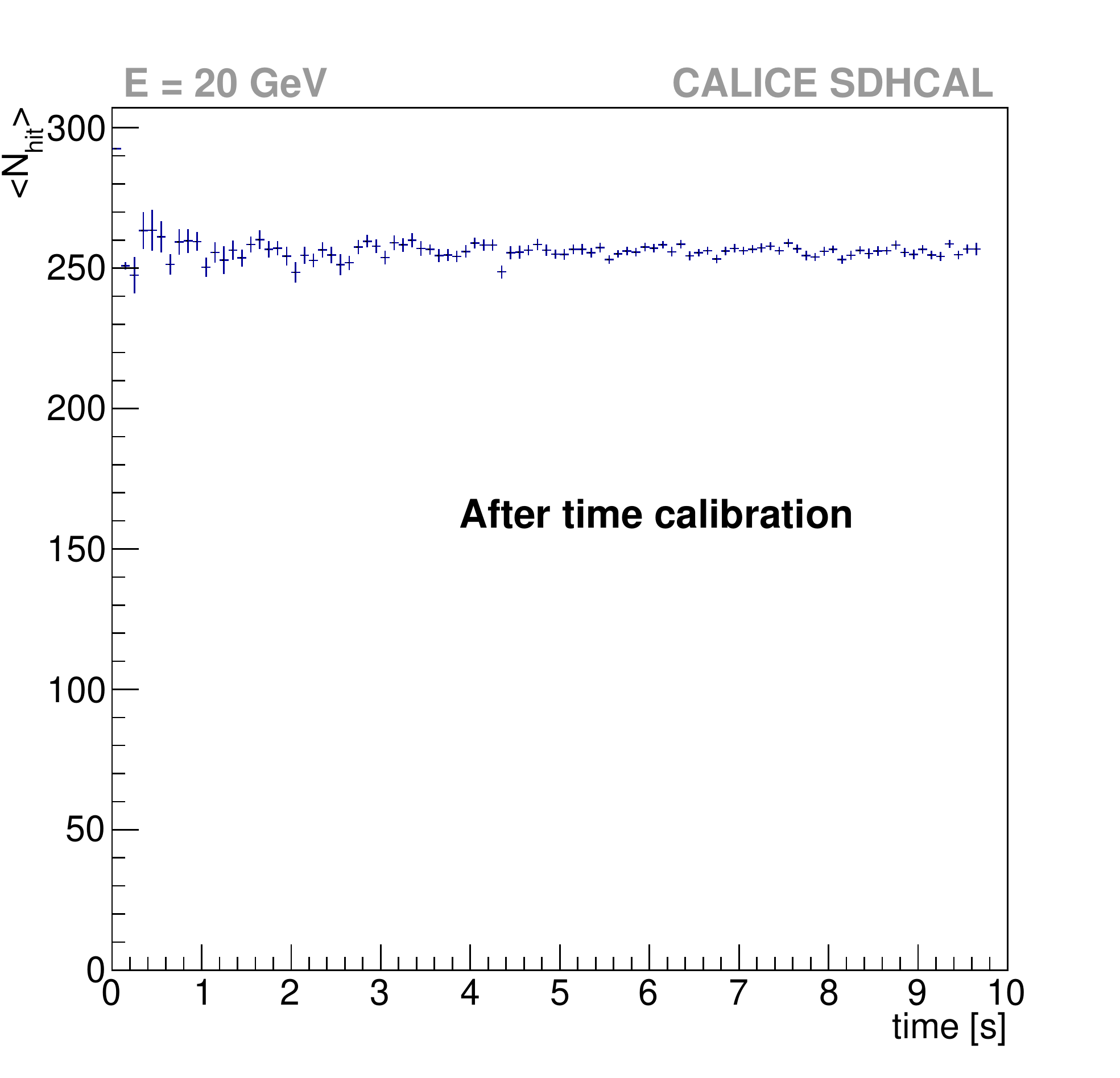}
    \caption{ }
  \end{subfigure}
  \caption{Average number of hits as a function of spill time for a 20~GeV electron run before (a) and after (b) the time calibration. The red curve is the result of the fit.}
  \label{fig.timefit}
\end{figure}

%\graphicspath{{figs/results/}}
The same data event building procedure as described in section~\ref{sec:method} is used. Because no Cherenkov counter was used during the beam tests, a topological selection is needed to identify the particle type. Muon track events are rejected by requesting that the average number of hits per layer is a few sigmas higher than the average pad multiplicity value. More details concerning the selection can be found in \cite{energyReco}. Electromagnetic and hadronic shower selections contain few additional requirements which will be described in sections~\ref{subsec.elec} and \ref{subsec.pion}. During the data taking period the beam was set to have less than 1000 particles per spill (the SPS spill was around 9 seconds every 45 seconds in 2012). This was intended to ensure a stable and good detection efficiency of muons. However with hadronic and electromagnetic showers, a decrease in the number of hits in the SDHCAL prototype during the spill time (see Fig.~\ref{fig.timefit}(a)) has been observed.
This effect increases with the deposited charge in the glass and so with the shower energy. This behavior is also more pronounced with electromagnetic showers due to their compactness. In the glass (whose electric bulk resistivity is around $10^{12}~\Omega \cdot$m), it takes time to absorb the electrons and the ions produced  during the avalanche. 
It was measured that SDHCAL GRPCs become less efficient at a rate exceeding 100~Hz/cm$^2$~\cite{yacine}. The reduction of the number of hits associated to events in the same run during the spill is higher for second and third thresholds that are triggered by higher deposited charge. To correct for this behaviour, the number of hits for each threshold and for each run is fitted with a polynomial function of the time measured with respect to the starting time of the spill as shown in Fig.~\ref{fig.timefit}(a). The corrected number of hits for threshold $i$ ($N_i^{corr}$) is then defined as: 
\begin{equation}
  N_i^{corr}=N_i-\sum_{j=1}^d{p_jt^j}\ \ ;\ \  i=1,2,3.
  \label{eq.timecorrection}
\end{equation}
where $d$ is the degree of the polynomial correction and $t$ is the relative time in seconds with respect to the starting time of the spill. For hadronic showers $d=1$ was found to fairly correct the number of hits for the three thresholds while for electromagnetic showers, due to denser charge deposits, $d=3$ was needed. Figure~\ref{fig.timefit}(b) shows the average number of hits as a function of spill time for a 20~GeV electron run after the calibration. In the following, the number of hits will refer to the corrected number of hits for each threshold and for the total number of hits defined by :
\begin{equation}
  N_{hit} = \sum_{i=1}^{3}{N_i^{corr}}
\end{equation}
%%%%%%%%%%%%%%%%%%%%%%%%%%%%%%%%%%%%%%%%%%%%%%%%%
\subsection{Electromagnetic shower results}
\label{subsec.elec}
\begin{figure}[t]
  \begin{center}
    \includegraphics[width=0.45\textwidth]{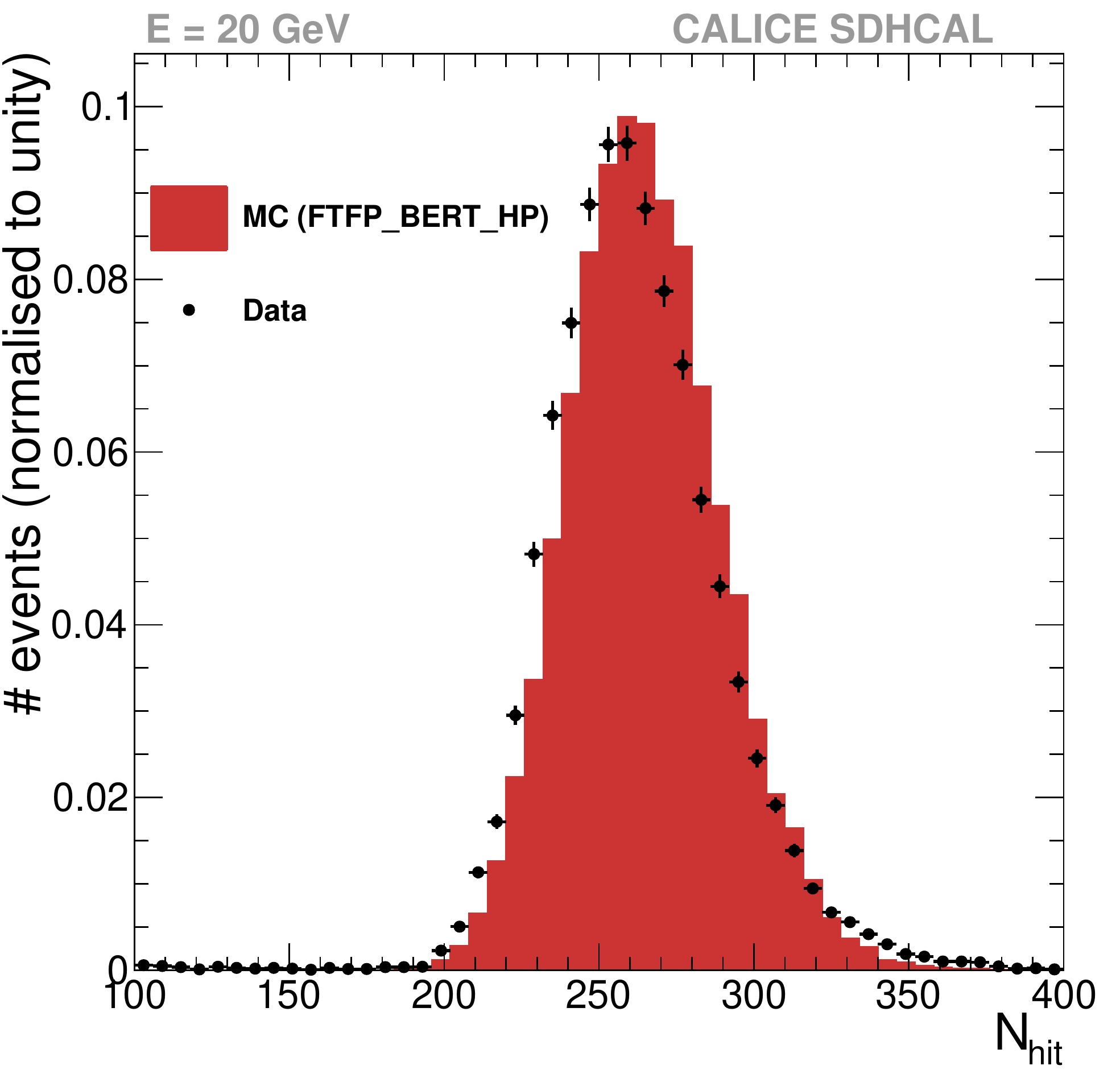}
    \includegraphics[width=0.45\textwidth]{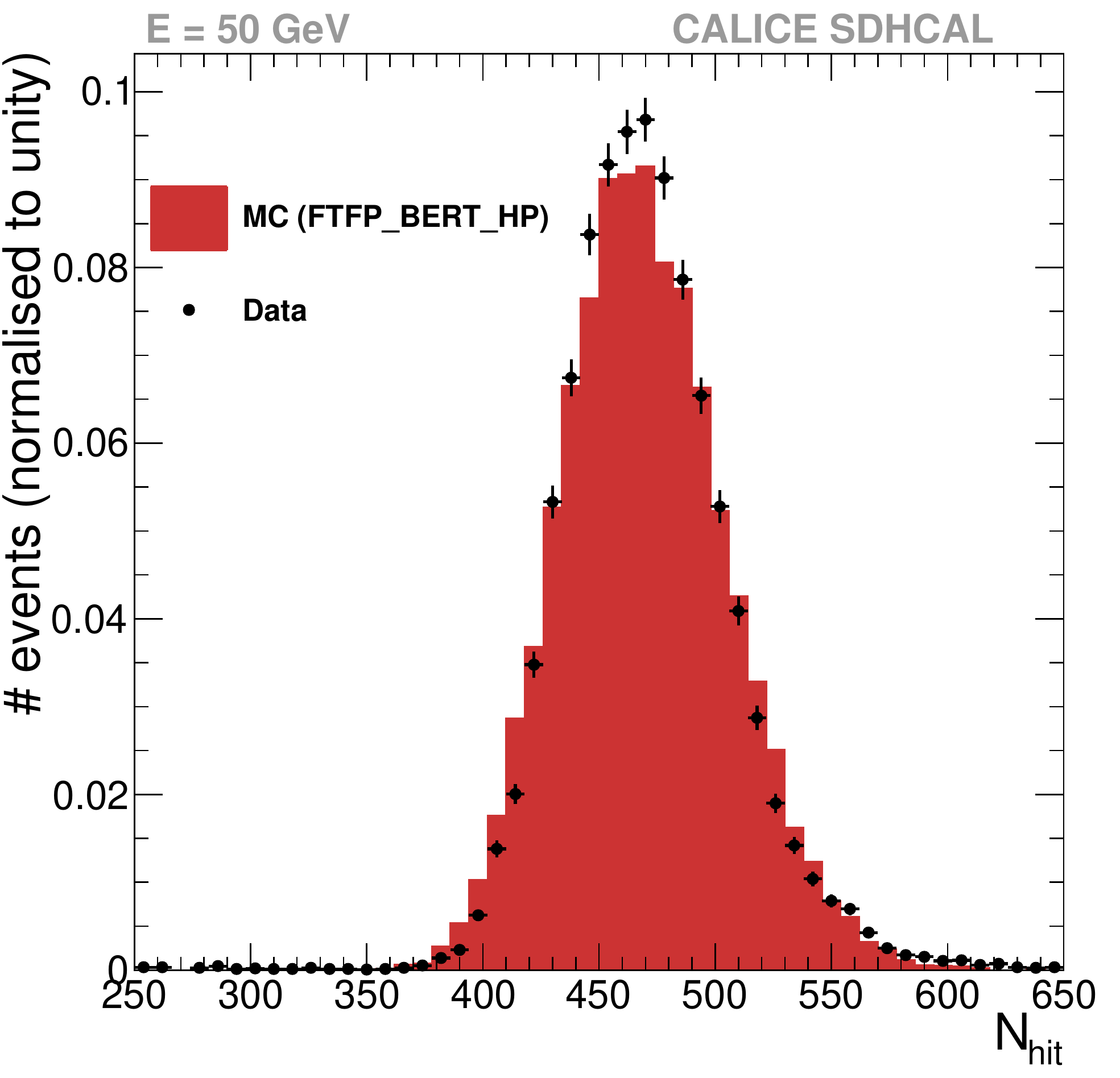}
    \caption{Number of hits distribution for 20 (left), and 50~GeV (right) electron runs. Data are represented by black circles and simulation by red filled histogram.}
    \label{fig.nhit_e-_dist}
  \end{center}
\end{figure}

The additional cuts applied to select electromagnetic showers are presented below:
\begin{enumerate}[~~1-]
\item The number of layers with at least one hit should be lower than 30 out of total of 48.
\item The number of reconstructed tracks using the Hough Transform technique as in \cite{sdhcalCalor} must be zero.
\item The first interaction layer should be located before the fifth layer of the detector. It is defined as the first layer with at least 4 hits and the same requirement for the three following layers.
\end{enumerate}
\begin{figure}[t]
  \begin{center}
    \includegraphics[width=0.49\textwidth]{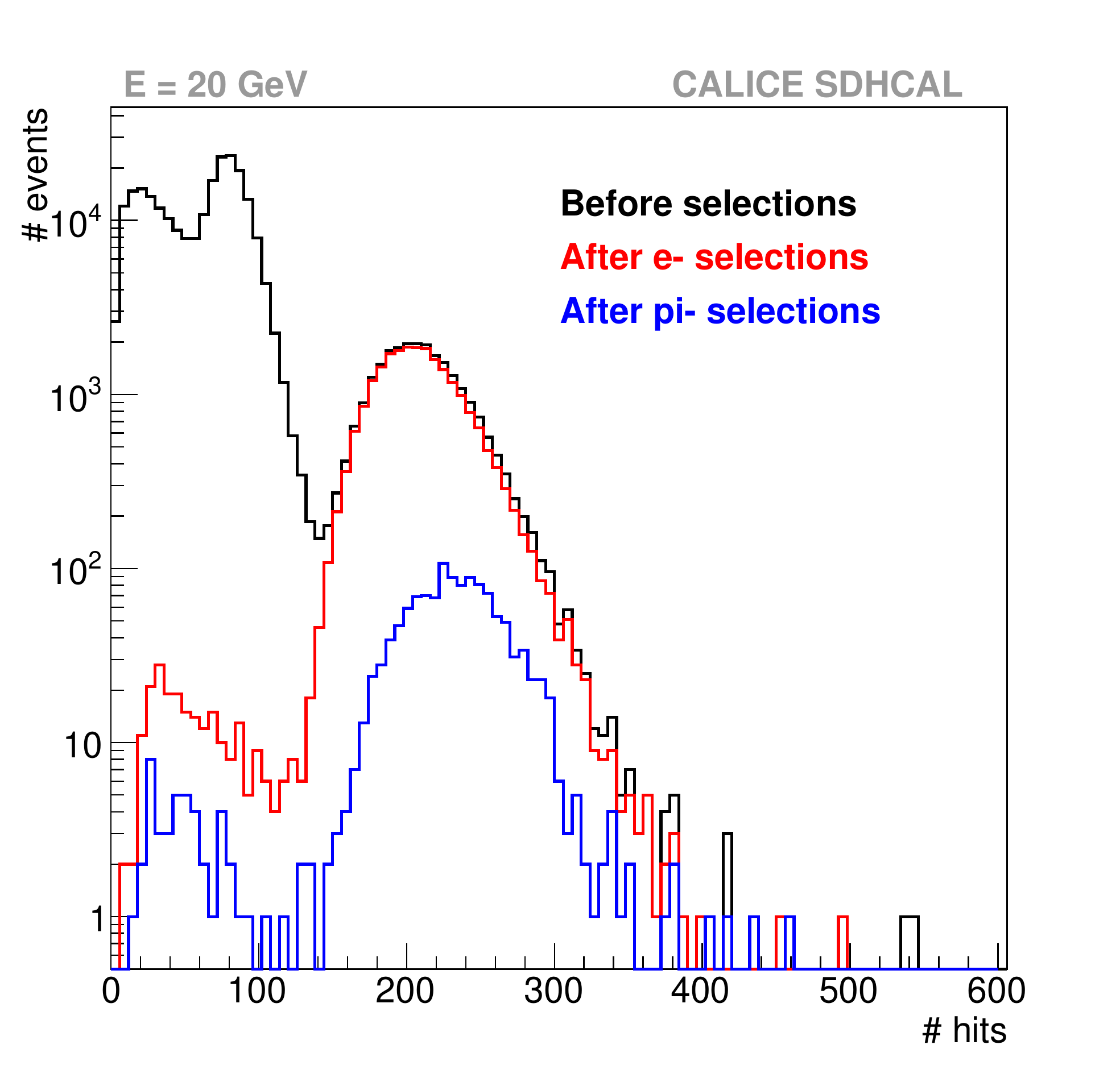}
    \includegraphics[width=0.49\textwidth]{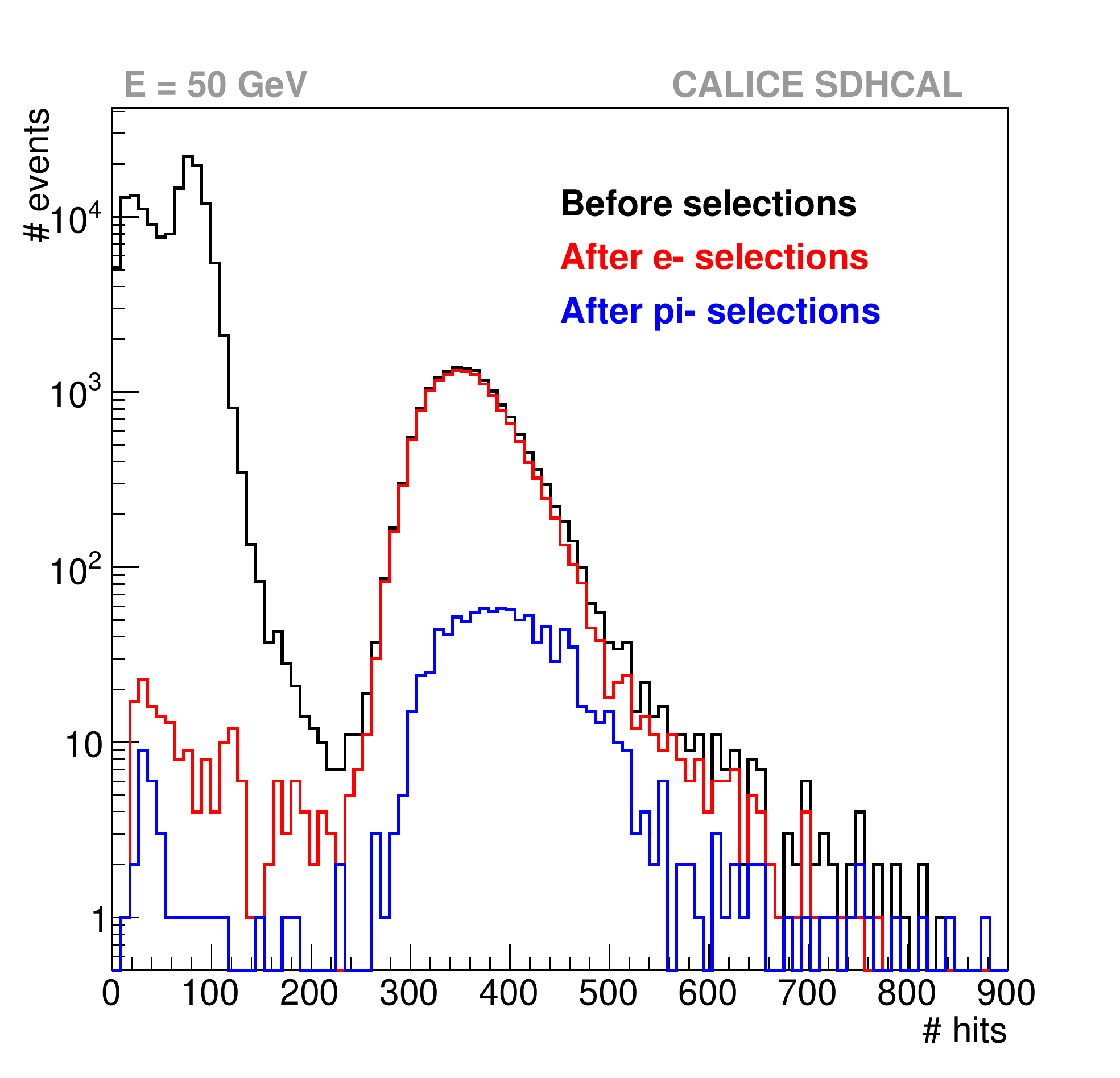}
    \caption{Distribution of number of hits without time correction for 20~GeV (left) and 50~GeV (right) electron runs. Black lines show the hit distributions before selection, red lines show the hit distributions after electron selection and blue lines show the hit distributions after pion selection.}
    \label{fig.e-selection}
  \end{center}
\end{figure}
Figure~\ref{fig.e-selection} shows the hit distributions for 20 and 50~GeV electron runs before and after the application of these selection criteria. The distributions of number of hits after pion selection (see section~\ref{subsec.pion}) are also shown. The selection effiency, determined using simulated samples of electromagnetic showers is higher than $99\%$ on the whole energy range ($[10;50]$~GeV). Figure~\ref{fig.nhit_e-_dist} shows hit distributions for 20 and 50~GeV electron runs for both data and simulation. As it was explained in sections~\ref{sec:method}~and~\ref{sec.parsummary}, the parameter $d_{cut}$ is tuned to reproduce, in the simulation, the number of hits of electromagnetic shower data.
\begin{figure}[t]
  \begin{center}
    \includegraphics[width=0.45\textwidth]{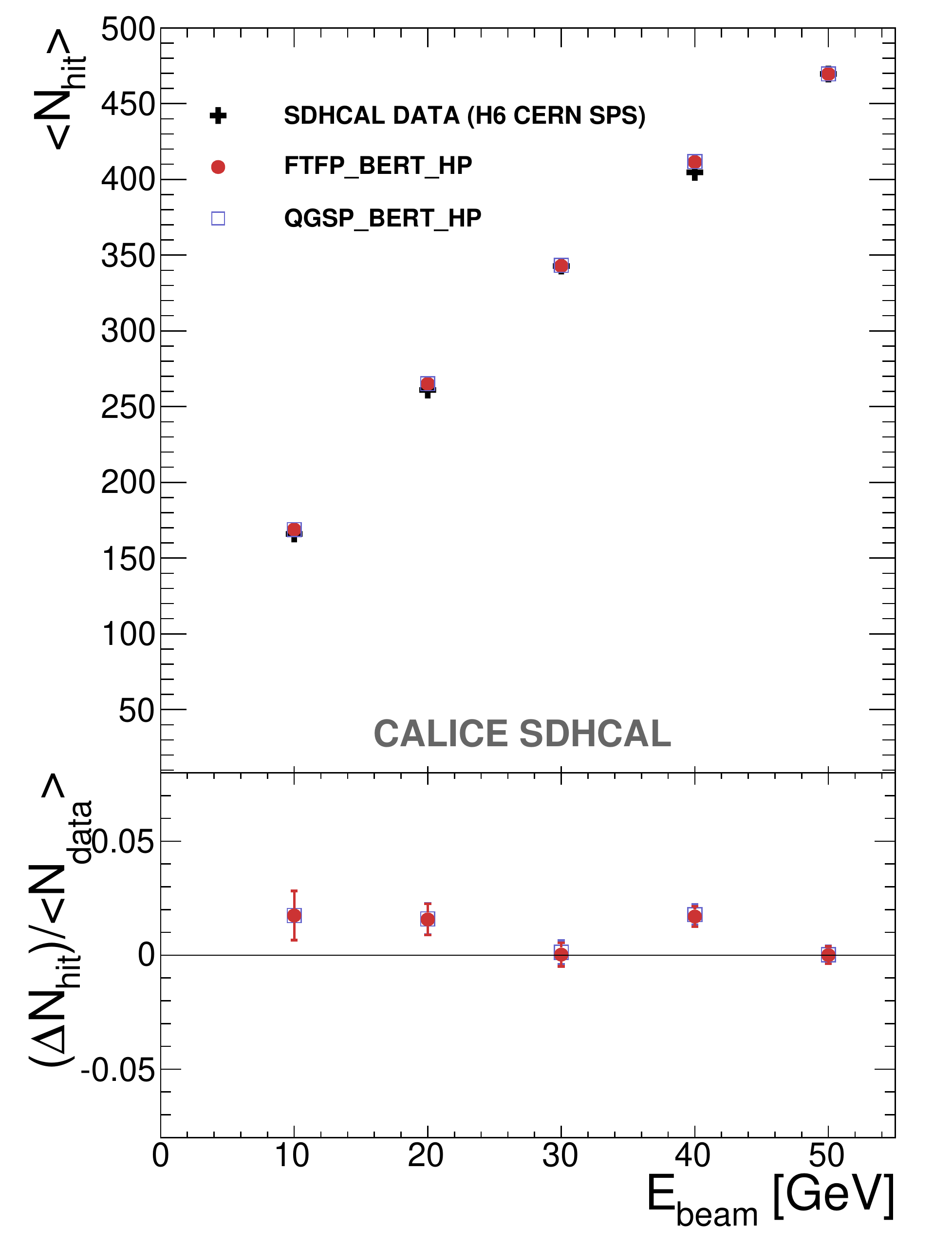}
    \caption{Average number of hits as a function of the beam energy for electron runs. Data are represented by black crosses, simulations are represented by red circles and open blue squares for FTFP\_BERT\_HP and QGSP\_BERT\_HP physics lists respectively. Relative deviations $\Delta N_{hit}/\langle N_{data} \rangle$ are also presented (see text for details).}
    \label{fig.nhit_e-_vs_ebeam}
  \end{center}
\end{figure}
Figure~\ref{fig.nhit_e-_vs_ebeam} shows the mean value of number of hits $\langle\textmd{N}_{hit}\rangle $ for both data and simulation and the relative deviation (defined as $\frac{\langle \textmd{N}_{\textmd{hit}}^{sim} \rangle - \langle \textmd{N}_{\textmd{hit}}^{data} \rangle}{ \langle \textmd{N}_{\textmd{hit}}^{data} \rangle}$) as a function of beam energy. The results obtained with the two physics lists are in agreement within statistical errors. This is expected as they both use the same model to simulate electromagnetic showers. The agreement between data and both simulation physics lists is satisfactory. The relative deviations are below 2$\%$ in the considered energy range. These results confirm the digitizer method and the chosen parametrization.
%%%%%%%%%%%%%%%%%%%%%%%%%%%%%%%%%%%%%%%%%%%%%%%%%
\subsection{Hadronic shower results}
\label{subsec.pion}
\begin{figure}[!ht]
  \begin{center}
    \includegraphics[width=0.49\textwidth]{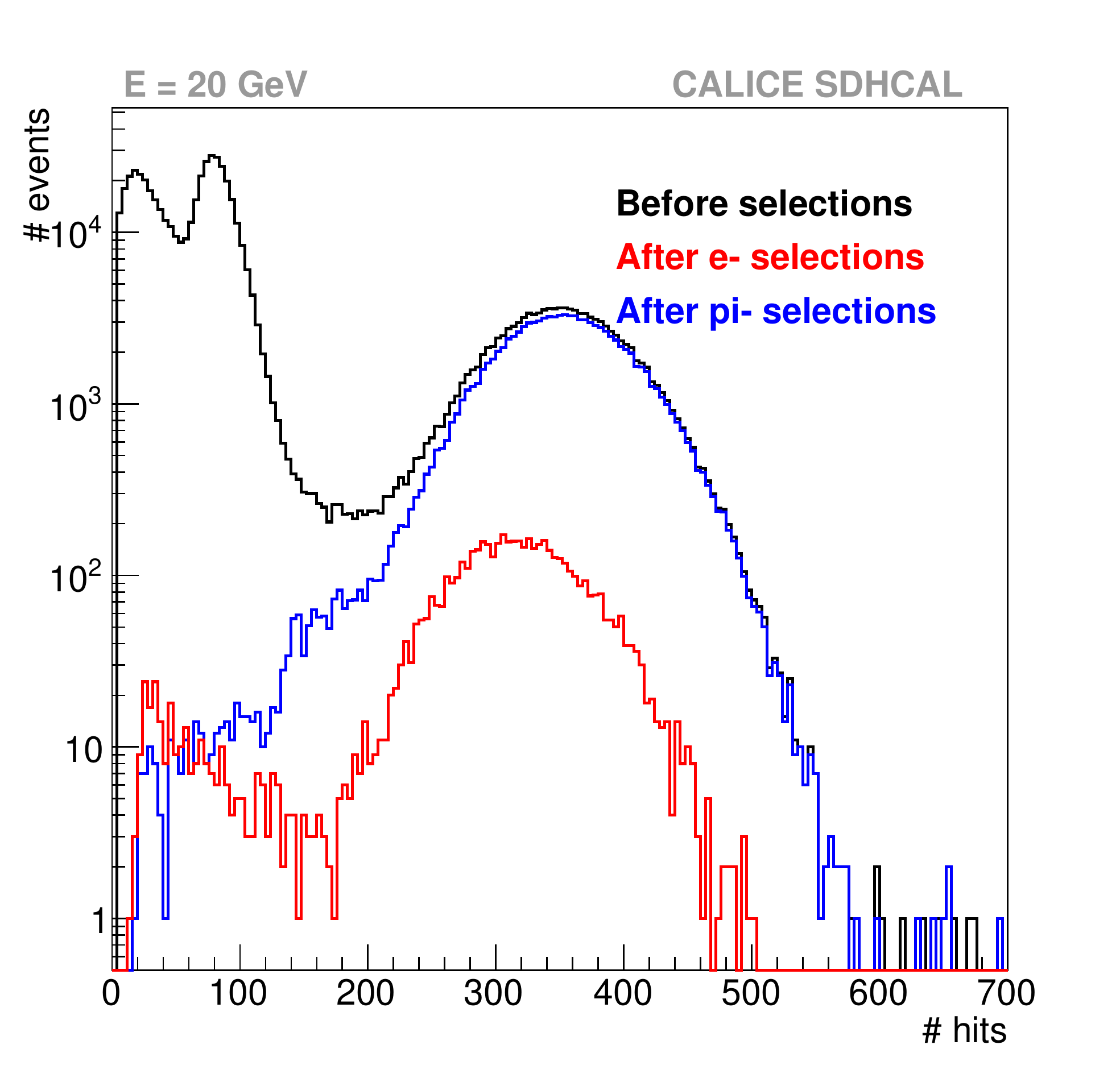}
    \includegraphics[width=0.49\textwidth]{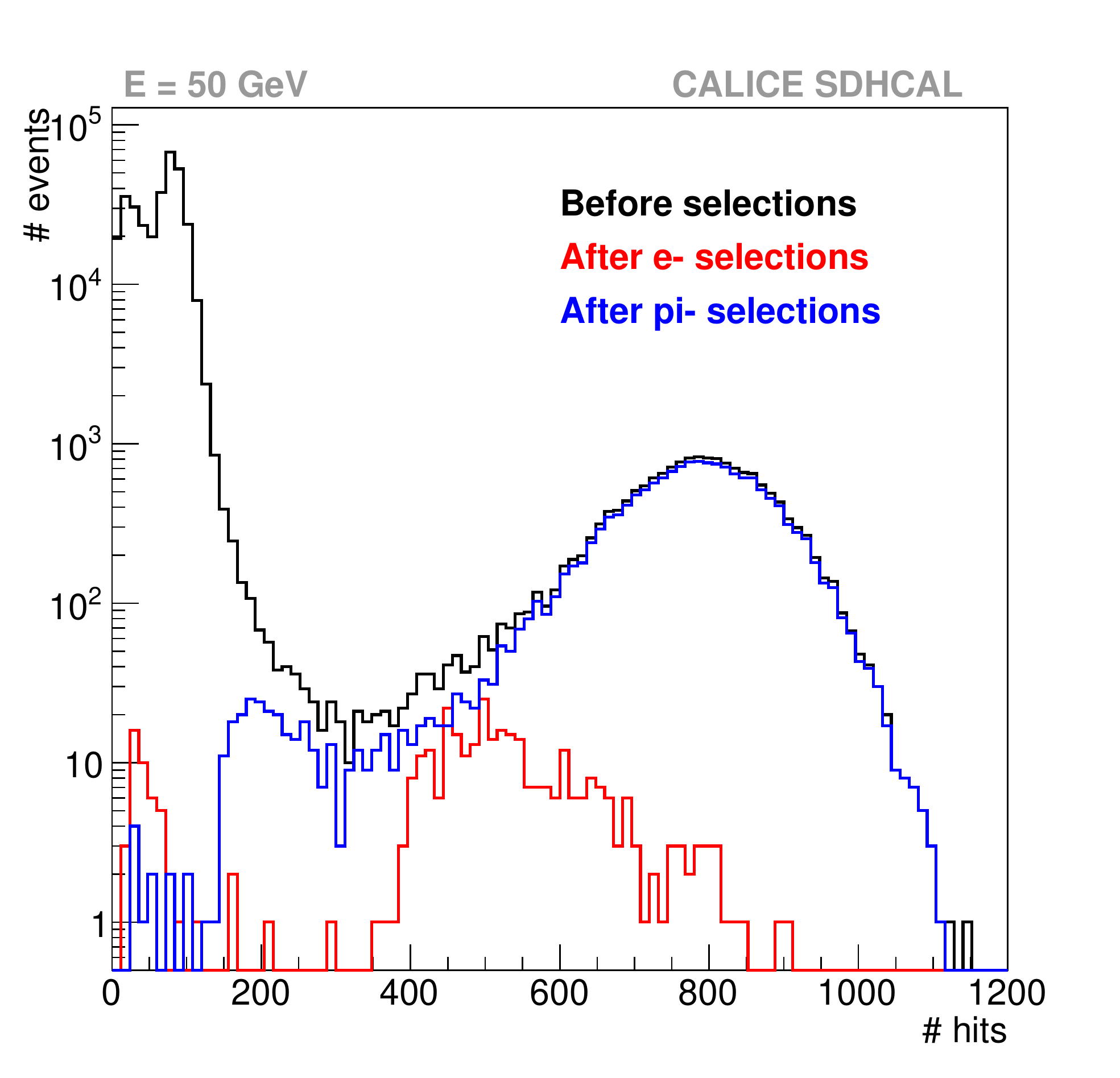}
    \caption{Distribution of number of hits without time correction for 20~GeV (left) and 50~GeV (right) pion runs. Black lines show the hit distributions before selection, red lines show the hit distributions after electron selection.}
    \label{fig.pi-selection}
  \end{center}
\end{figure}
\begin{figure}[t]
  \begin{center}
    \includegraphics[width=0.35\textwidth]{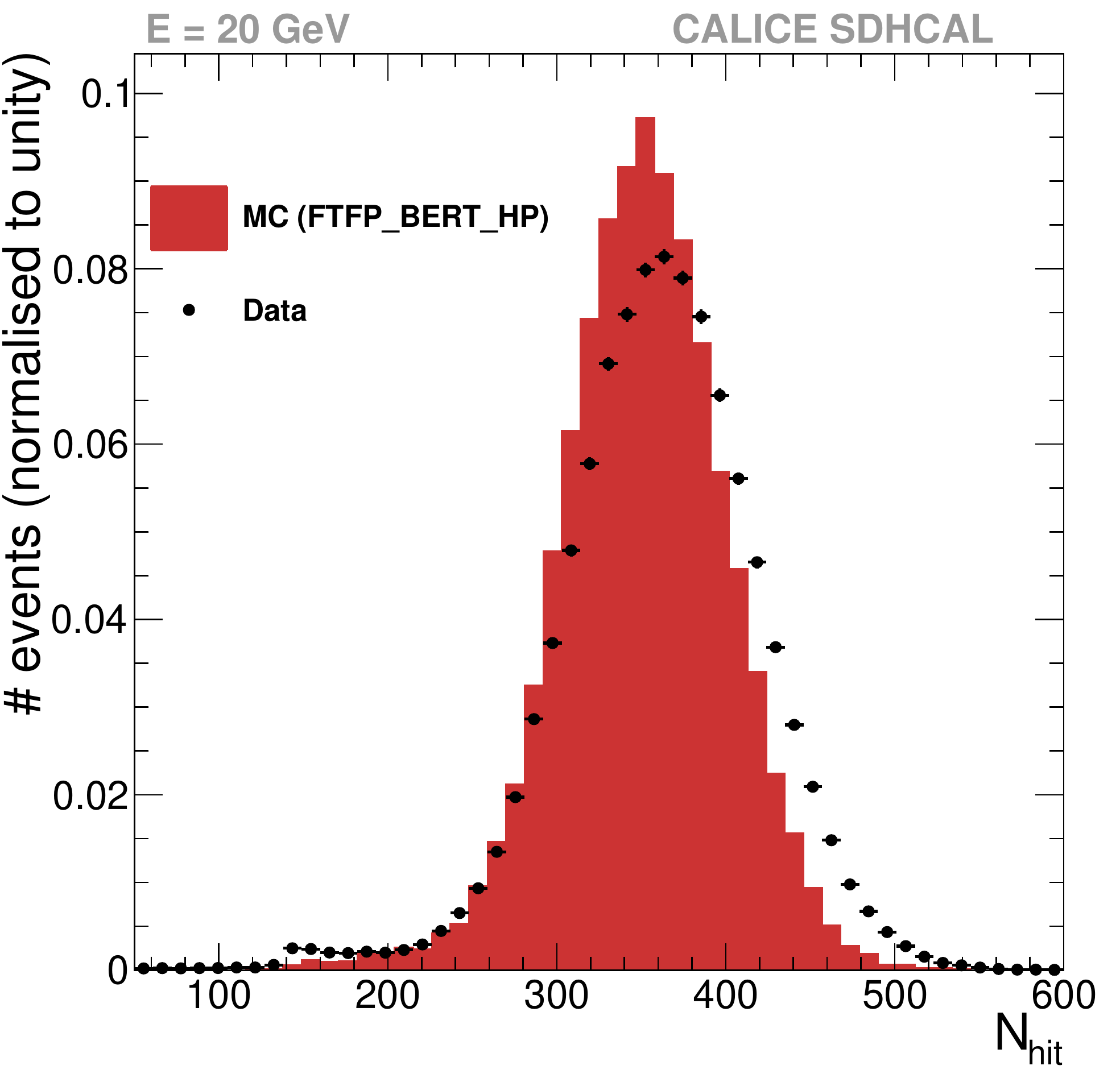}
    \includegraphics[width=0.35\textwidth]{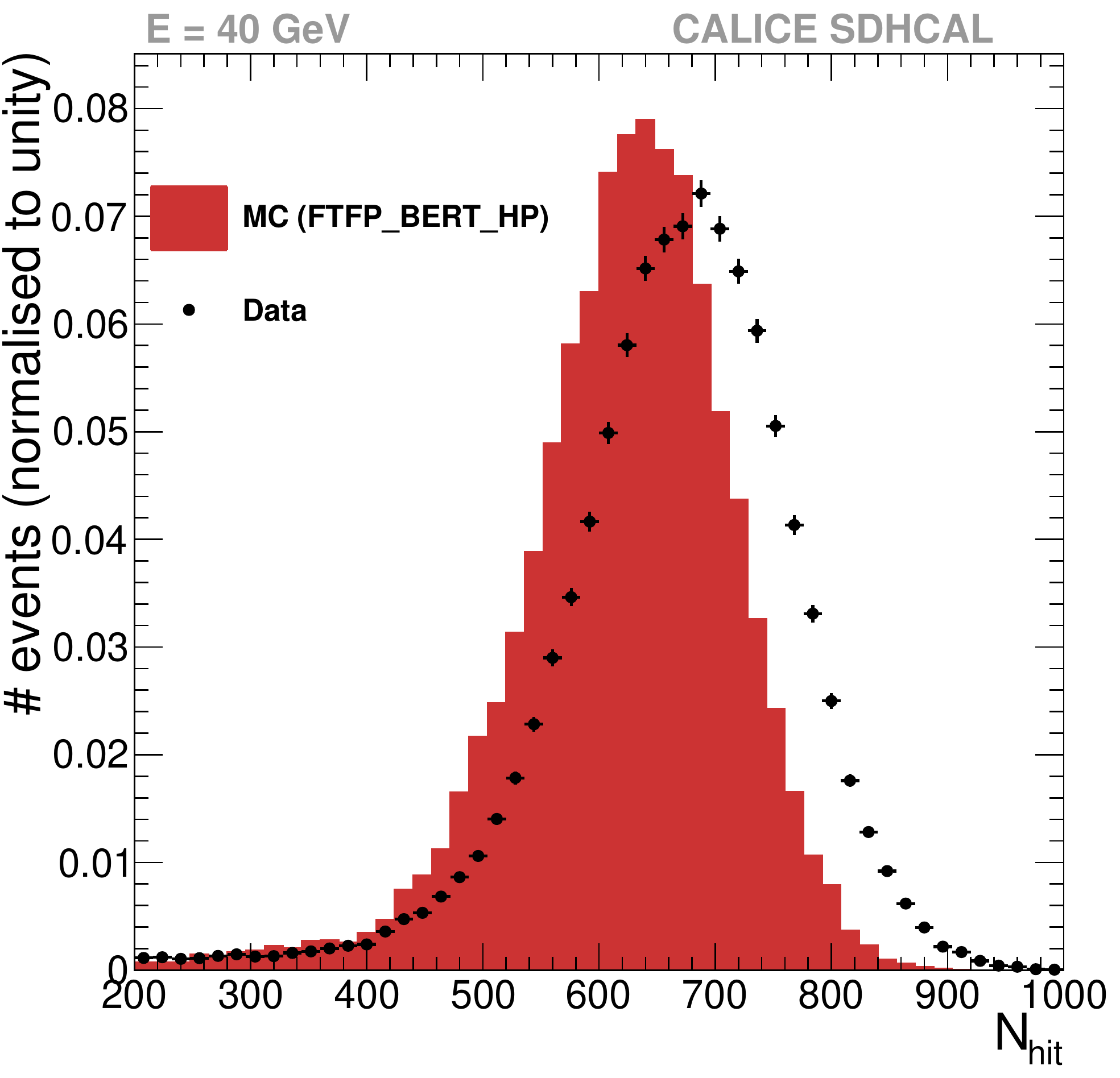}
    \includegraphics[width=0.35\textwidth]{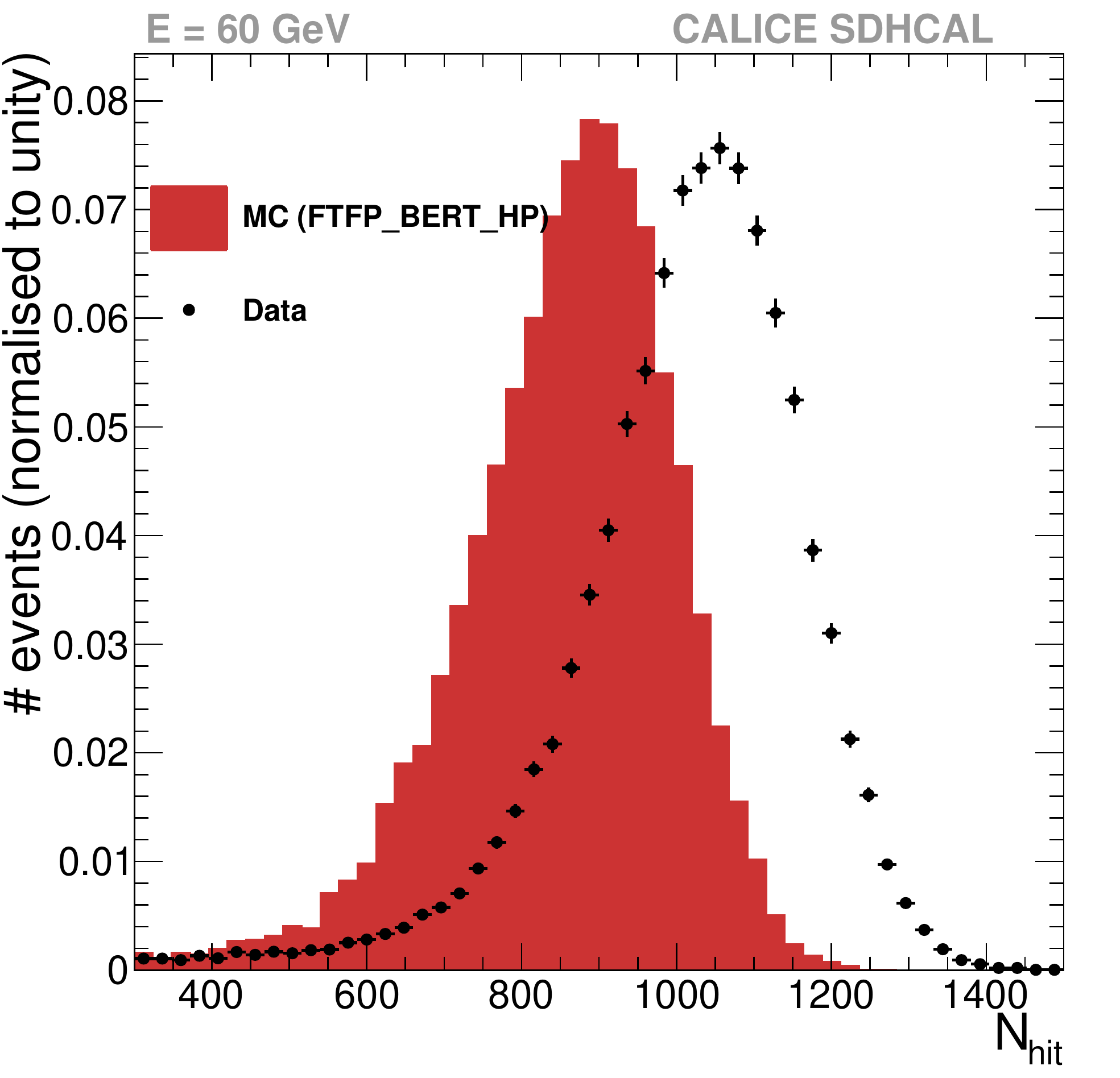}
    \includegraphics[width=0.35\textwidth]{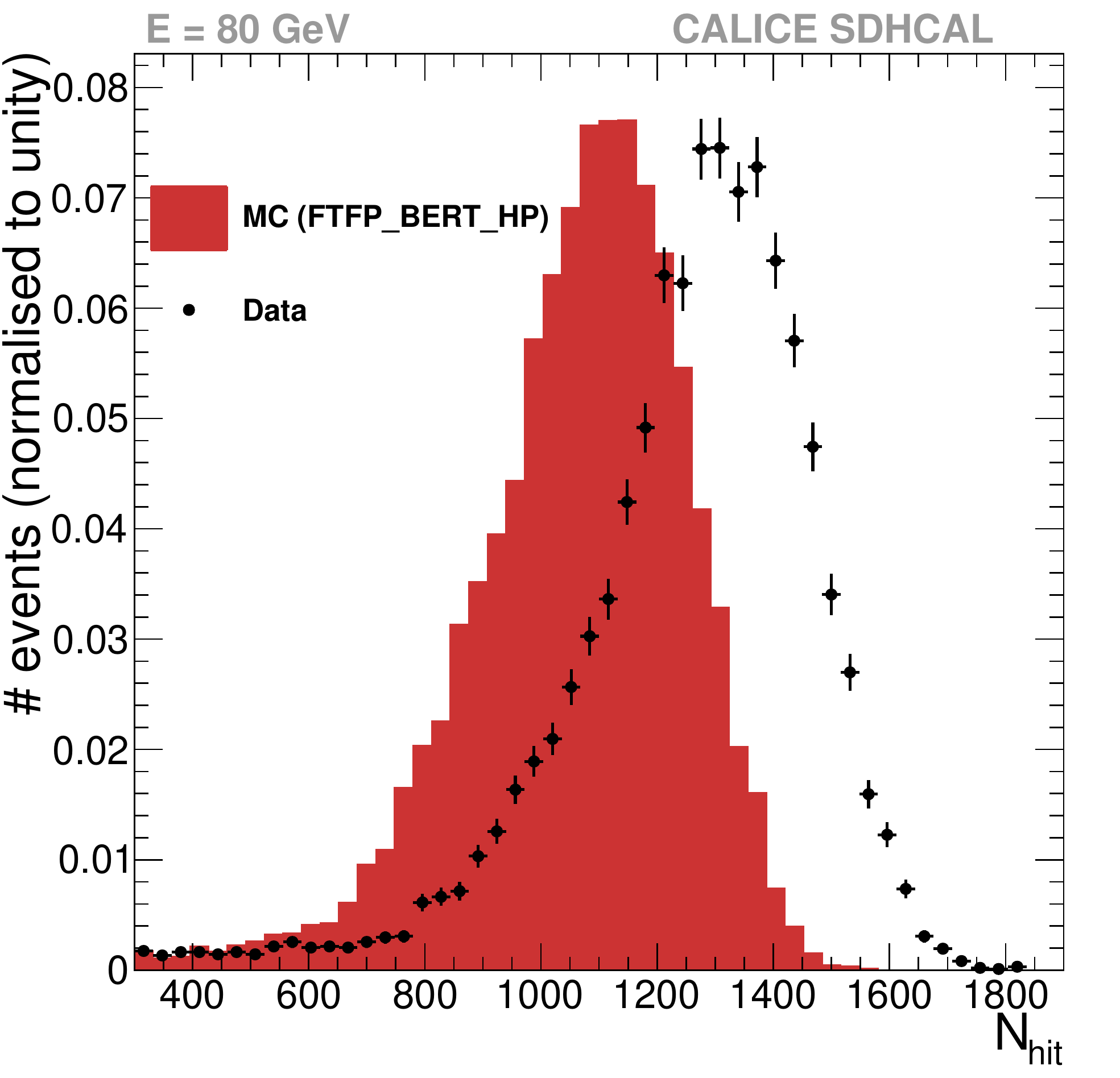}
    \caption{Number of hits distribution for 20, 40, 60 and 80~GeV pion runs. Data are represented by black circles and simulation by red filled histogram.}
    \label{fig.nhit_pi-_dist}
  \end{center}
\end{figure}
\begin{figure}[t]
  \centering 
  \includegraphics[width=0.4\textwidth]{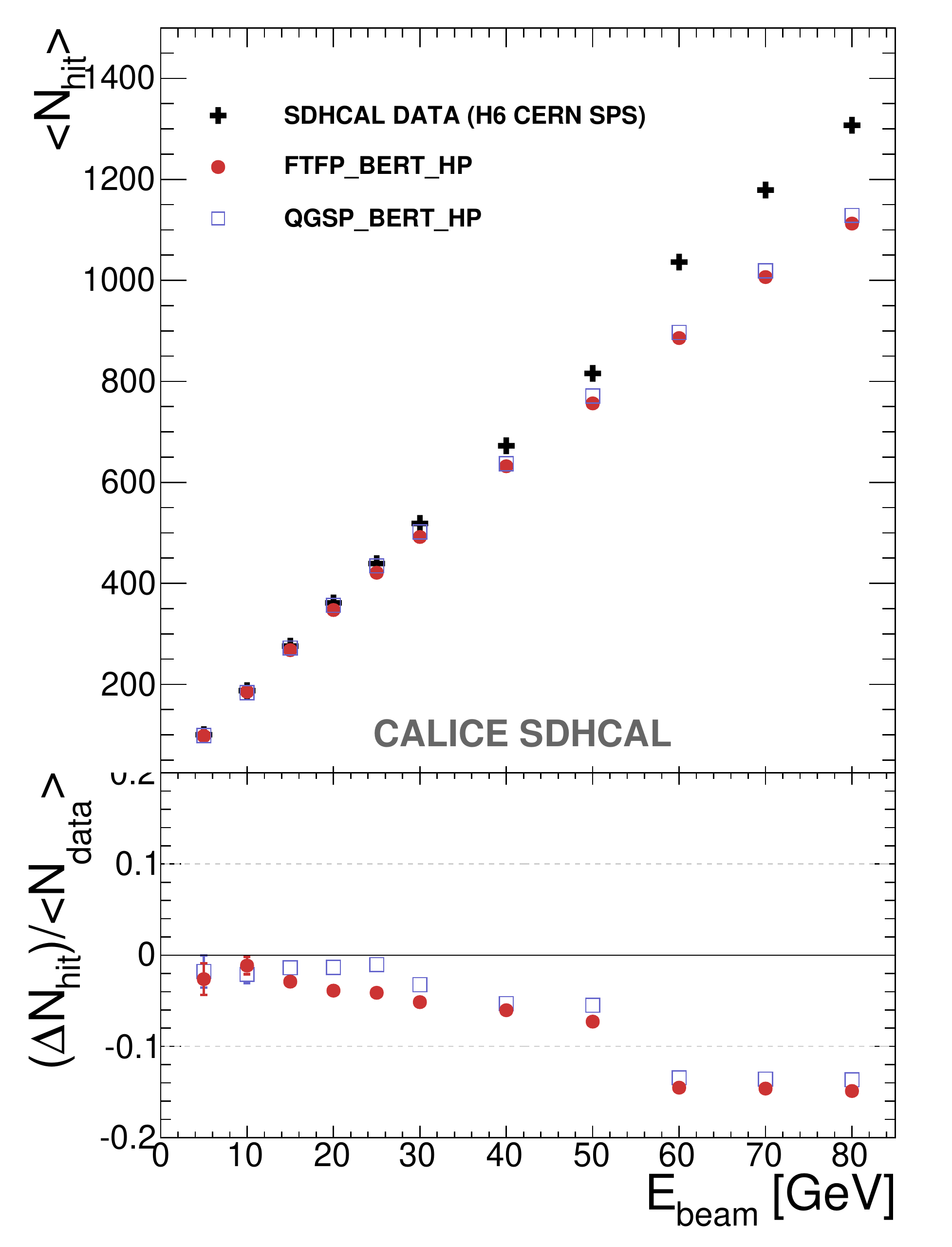}
  \caption{Average number of hits as a function of the beam energy for pion runs. Data are represented by black crosses, simulations are represented by red circles and open blue squares for FTFP\_BERT\_HP and QGSP\_BERT\_HP physics lists respectively. Relative deviations $\Delta N_{hit}/\langle N_{data} \rangle$ are also presented (see text for details).}
  \label{fig.nhit_pi-_vs_ebeam}
\end{figure}

To remove electromagnetic showers from the data samples at least one of the three following conditions must be satisfied: 
\begin{enumerate}[~~1-]
\item At least one track using the Hough Transform algorithm must be found.
\item The shower starting layer is located after the fifth layer. 
\item The number of fired layers is greater than 30. 
\end{enumerate}
Figure~\ref{fig.pi-selection} shows the hit distributions for 20 and 50~GeV pion runs before and after the application of these selection criteria. The distributions of number of hits after the electromagnetic shower selection are also shown (see section~\ref{subsec.elec}).
As for electromagnetic showers, selection efficiency is calculated using simulated samples of hadronic showers. The efficiency is $51\%$ at 5~GeV, $86\%$ at 10~GeV and above $92\%$ for energies higher than 15~GeV.
Figure~\ref{fig.nhit_pi-_dist} shows distributions of hits from hadronic shower runs for 4 different beam energies for both data and simulated events. Figure~\ref{fig.nhit_pi-_vs_ebeam} presents the mean value of the number of hits and the relative deviation as a function of beam energy.

The agreement between data and simulation is within 5$\%$ up to 30~GeV and is significantly degraded at higher energies. Proton contamination of the H6 SPS beam line was suspected to be the reason for these differences. The ATLAS Collaboration measured the fraction of protons in H6 to be significant (up to 61$\%$ at 100~GeV) \cite{atlas}. Since the proton interaction length is slightly lower than the pion's one~\cite{pion-proton}, the longitudinal leakage should be lower for proton than for pion showers. This leads to a slightly higher number of hits for proton than for pion showers. 
\begin{figure}[t]
  \centering 
  \begin{subfigure}[b]{.4\textwidth}
    \includegraphics[width=1.0\textwidth]{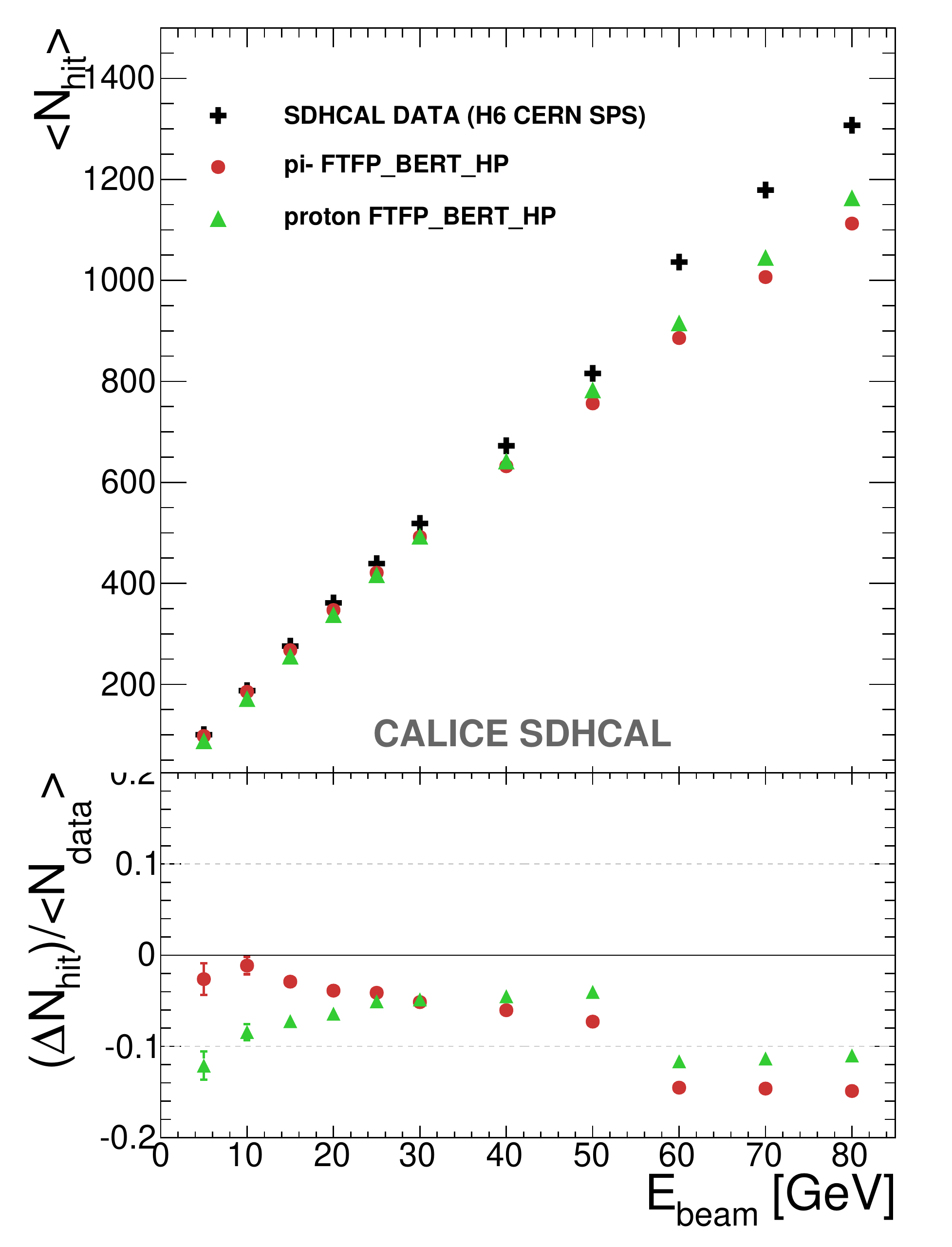}
    \caption{ }
  \end{subfigure}
  \begin{subfigure}[b]{.4\textwidth}
    \includegraphics[width=1.0\textwidth]{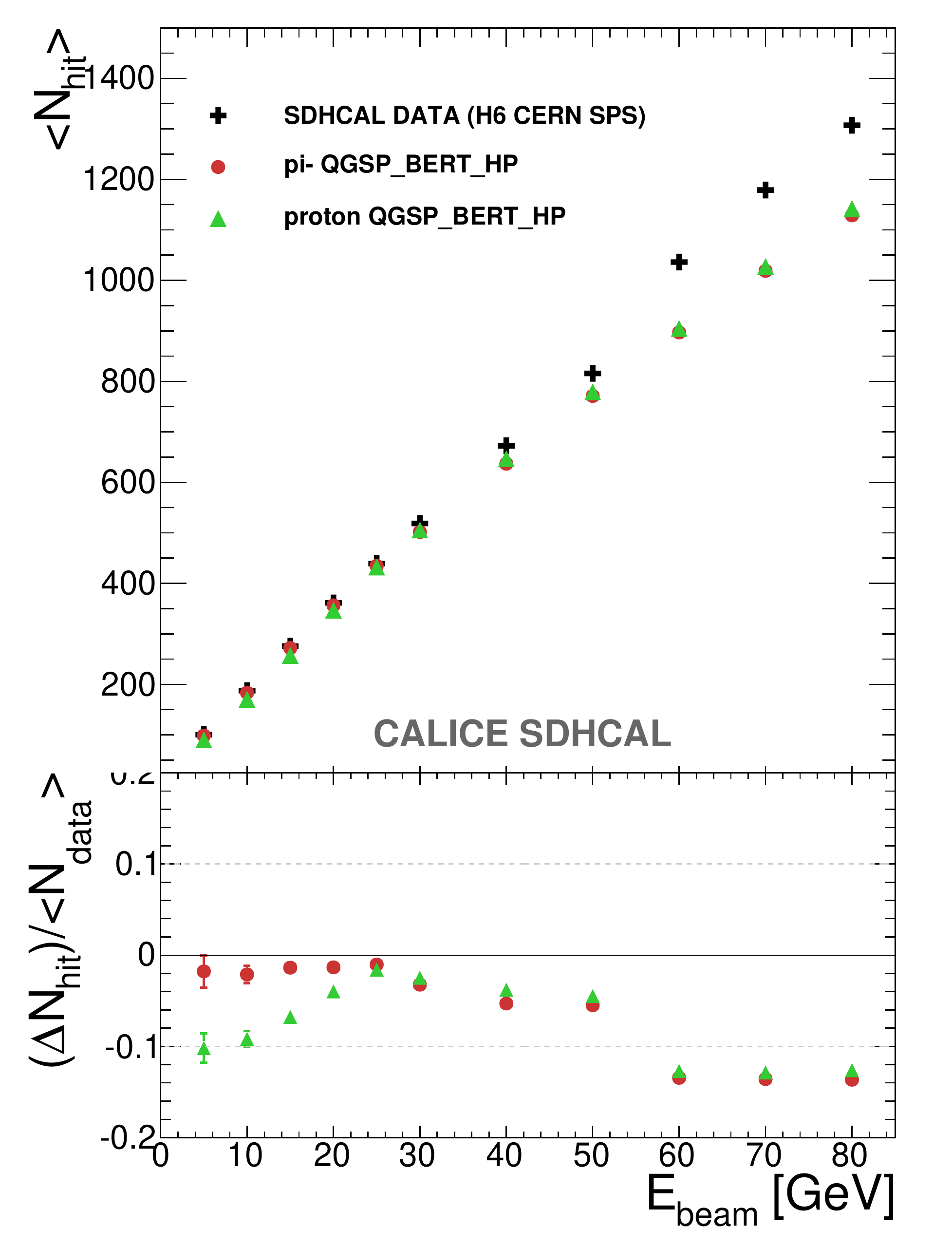}
    \caption{ }
  \end{subfigure}
  \caption{Average number of hits as a function of the beam energy for data, pion simulations and proton simulations. Comparison is shown for both physic lists FTFP\_BERT\_HP~(a) and QGSP\_BERT\_HP~(b). Data are represented by black crosses, pion simulations is represented by red circles and proton simulations by green triangles. Relative deviations $\Delta N_{hit}/\langle N_{data} \rangle$ are also presented (see text for details).}
  \label{fig.nhit_vs_ebeam_proton}
\end{figure}
Figure~\ref{fig.nhit_vs_ebeam_proton} shows the mean number of hits as function of beam energy for data as well as for both pion and proton obtained with the simulation using two different physics lists. At high energy, the number of hits for simulated proton showers is slightly higher than that for the simulated pion showers for the FTFP\_BERT\_HP physics list. However the number of hits for simulated proton showers is still significantly lower than what is observed in the data. This indicates that proton contamination cannot explain the observed difference at high energy between data and simulation. 

\begin{figure}[t]
  \centering
  \includegraphics[width=0.4\textwidth]{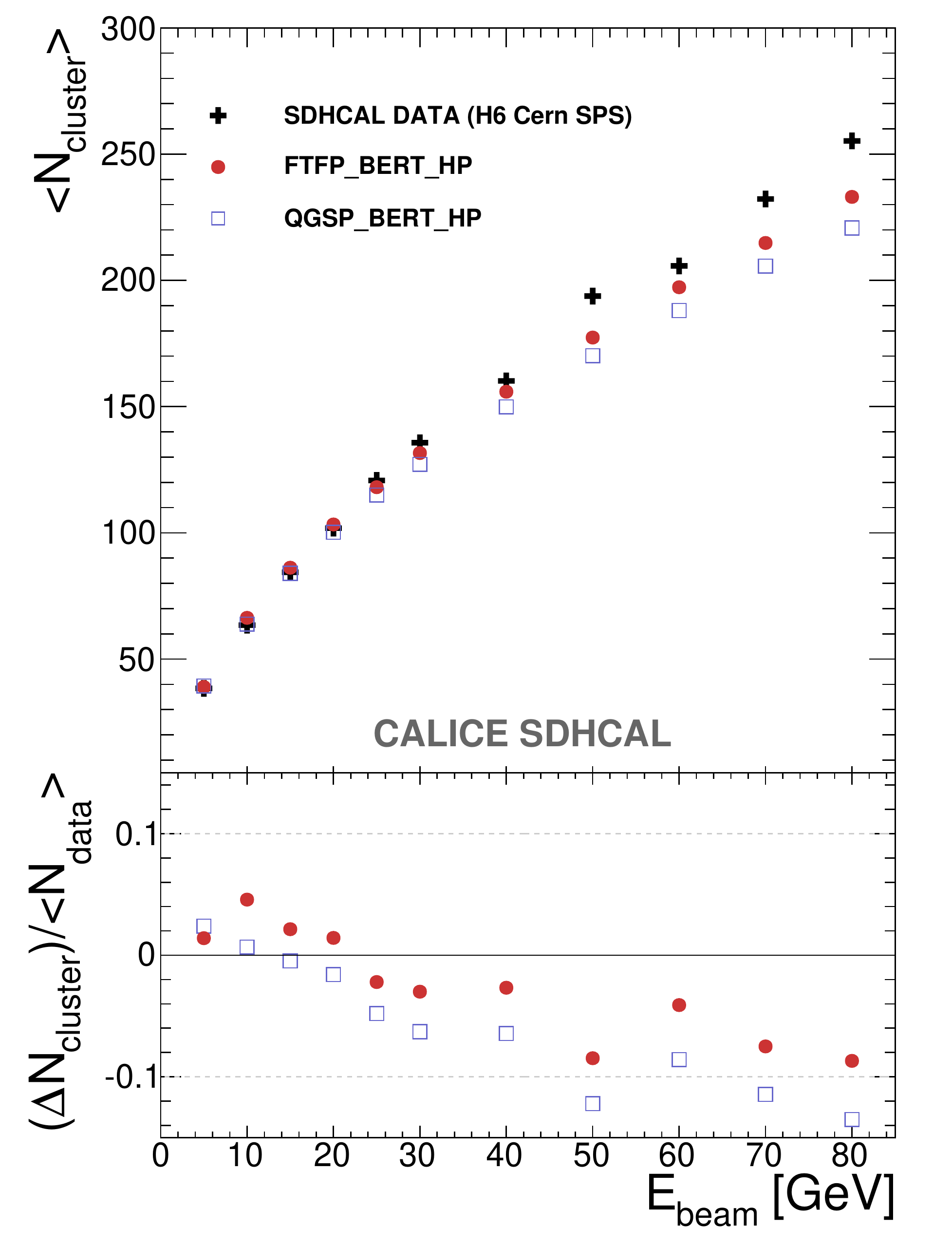}
  \caption{Average number of reconstructed clusters for pion runs as a function of the beam energy. Data are represented by black crosses, simulations are represented by red circles and open blue squares for FTFP\_BERT\_HP and QGSP\_BERT\_HP physics lists respectively. Relative deviations $\Delta N_{cluster}/\langle N_{data} \rangle$ are also presented (see text for details).}
  \label{fig.nclusters_vs_ebeam}
\end{figure}
We also suspected that the parameterization in the charge splitting procedure (Eq.~\ref{eq.ratio} in section~\ref{sec:method}) could be responsible for the disagreement between the data and the simulation for the number of hits. To validate or reject this hypothesis, the reconstructed number of clusters was studied. A cluster is defined as a group of fired pads (hits) that are in the same layer and sharing  an edge. Figure~\ref{fig.nclusters_vs_ebeam} presents the average number of reconstructed clusters as a function of beam energy. The relative deviations (defined as $\frac{\langle \textmd{N}_{\textmd{cluster}}^{sim} \rangle - \langle \textmd{N}_{\textmd{cluster}}^{data} \rangle}{ \langle \textmd{N}_{\textmd{cluster}}^{data} \rangle}$) are also shown. This figure shows a satisfactory agreement between data and simulation below 40~GeV. The differences at higher energy between data and simulation confirm those observed on the total number of hits.

\begin{figure}[t]
  \centering 
  \begin{subfigure}[b]{.3\textwidth}
    \includegraphics[width=1.0\textwidth]{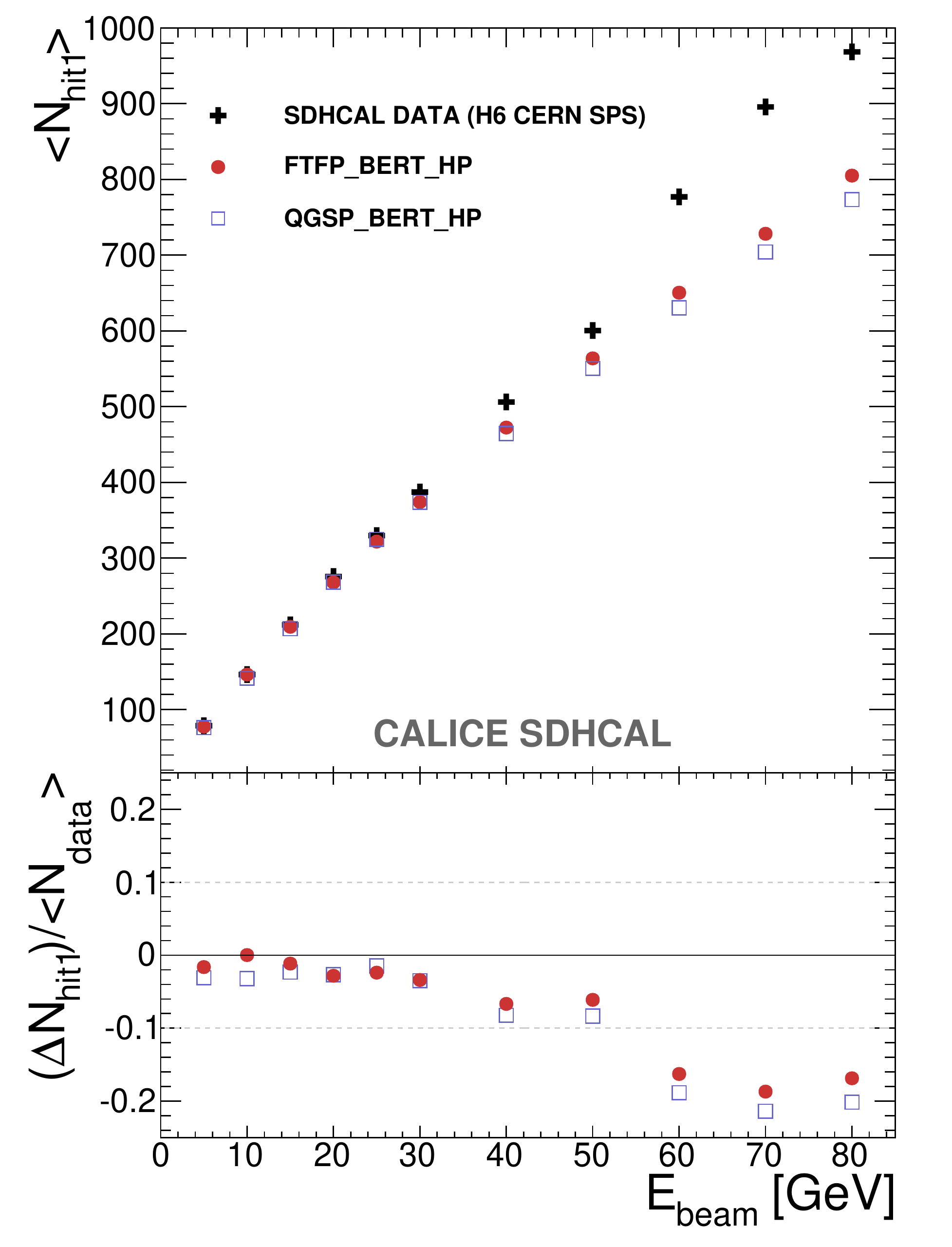}
    \caption{ }
  \end{subfigure}
  \begin{subfigure}[b]{.3\textwidth}
    \includegraphics[width=1.0\textwidth]{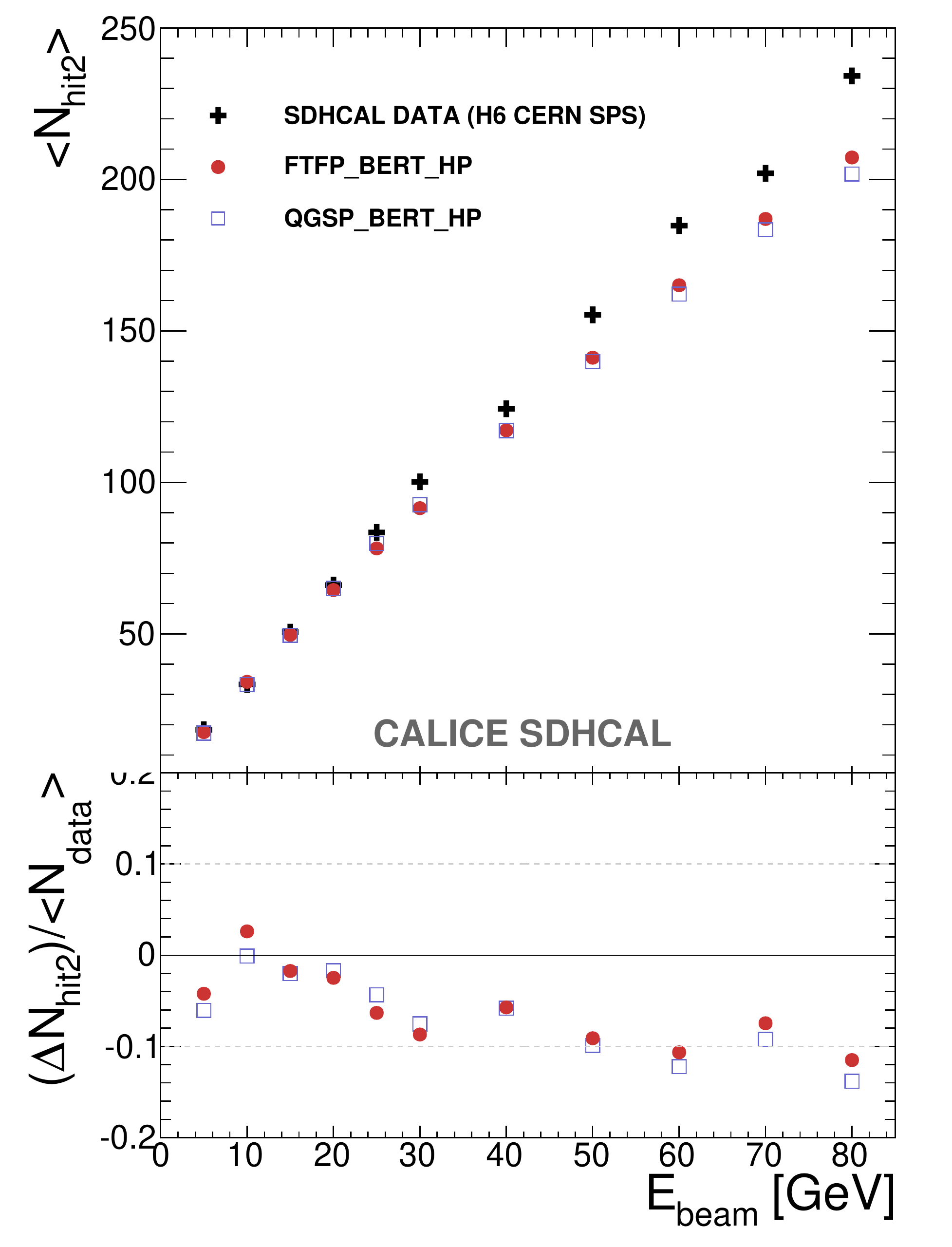}
    \caption{ }
  \end{subfigure}
  \begin{subfigure}[b]{.3\textwidth}
    \includegraphics[width=1.0\textwidth]{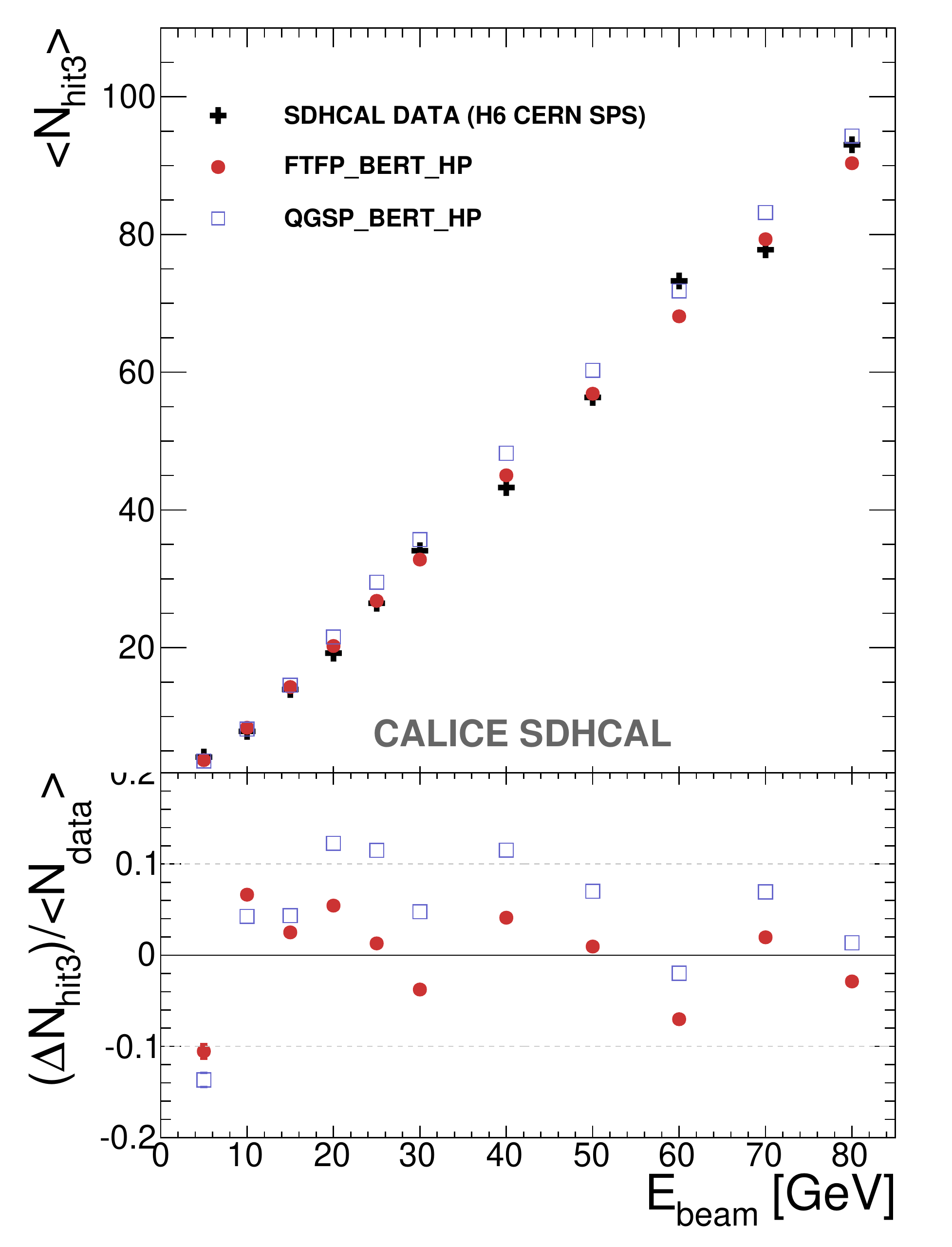}
    \caption{ }
  \end{subfigure}
  \caption{Average number of hits for threshold 1 (a), 2 (b) and 3 (c) for pion runs as a function of the beam energy. Data are represented by black crosses, simulations are represented by red circles and open blue squares for FTFP\_BERT\_HP and QGSP\_BERT\_HP physics lists respectively. Relative deviations $\Delta N_{hit,i}/\langle N_{data} \rangle$ are also presented (see text for details).}
  \label{fig.nhit_pi-_vs_energy_thr}
\end{figure}
Figure~\ref{fig.nhit_pi-_vs_energy_thr} shows the average number of hits for each threshold as a function of beam energy. The same behavior, observed for the total number of hits, is seen for the number of hits for the two first thresholds. The agreement between the data and the simulation degrades when the beam energy increases. The number of hits related to the third threshold observed in data is more or less well reproduced by the simulation. However, limited number of these hits makes difficult to draw a strong conclusion.

Some GEANT4 physics lists show satisfactory agreement with hadronic shower data obtained with other detector technologies. The Monte Carlo simulation was able to predict the hadronic shower response of the ATLAS-TileCal prototype within a few percents in a wide energy range (20~:~350~GeV) \cite{atlas}. The agreement between data and the simulated hadronic shower response in the CALICE-AHCAL prototype was also found to be satisfactory \cite{ahcal-geant}. However, the CALICE-AHCAL simulated response was higher than that in data above 30~GeV whereas an opposite behavior is observed within the SDHCAL prototype (Fig.~\ref{fig.nhit_pi-_vs_ebeam}). Nevertheless, for the CALICE-AHCAL prototype as well as for the ATLAS-TileCal prototype, the deposited energy was measured (analog readout) while the SDHCAL response is defined by the number of hits. Moreover, the transversal segmentation in ATLAS-TileCal ($\Delta \phi \times \Delta \eta \geq 0.1 \times 0.1$) and in CALICE-AHCAL ($\geq3\times3$~cm$^2$) is not as fine as in SDHCAL ($1\time1$~cm$^2$). This may explain why the number of hits (above 50~GeV) in the simulation was lower than that in the SDHCAL data while the agreement between data and simulation was better for the ATLAS-TileCal and CALICE-AHCAL prototype.

The radial shower profile was also studied using the CALICE-AHCAL prototype in \cite{ahcal-geant}. The conclusion of this study was that GEANT4 physics lists underestimate the radial extent of hadronic showers. The radial shower profile is also studied in the SDHCAL prototype. To compute this profile, the shower main thrust is estimated using a straight line fit of the unweighted shower hit positions. Then, the intersection of the main axis and each layer is used to locate the shower barycentre in each layer. Hits are then counted in 1~cm thick rings, using the center of pads, around the barycentre position. The number of hits for data in each ring is corrected with the spill time as it was done for shower number of hits. This correction is much more important in the core of the showers than in its periphery.
\begin{figure}[!h]
  \includegraphics[width=.45\textwidth]{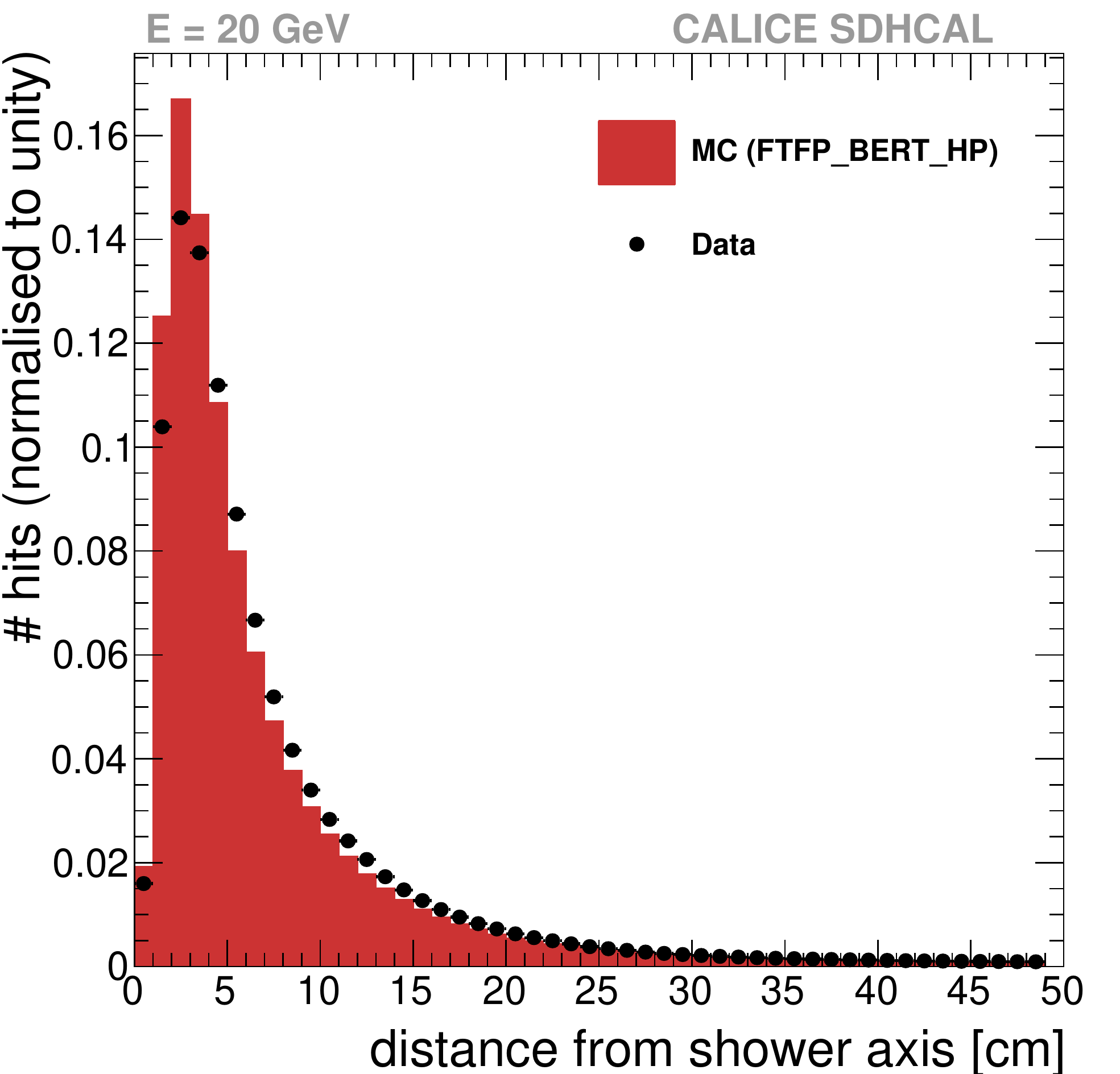}
  \includegraphics[width=.45\textwidth]{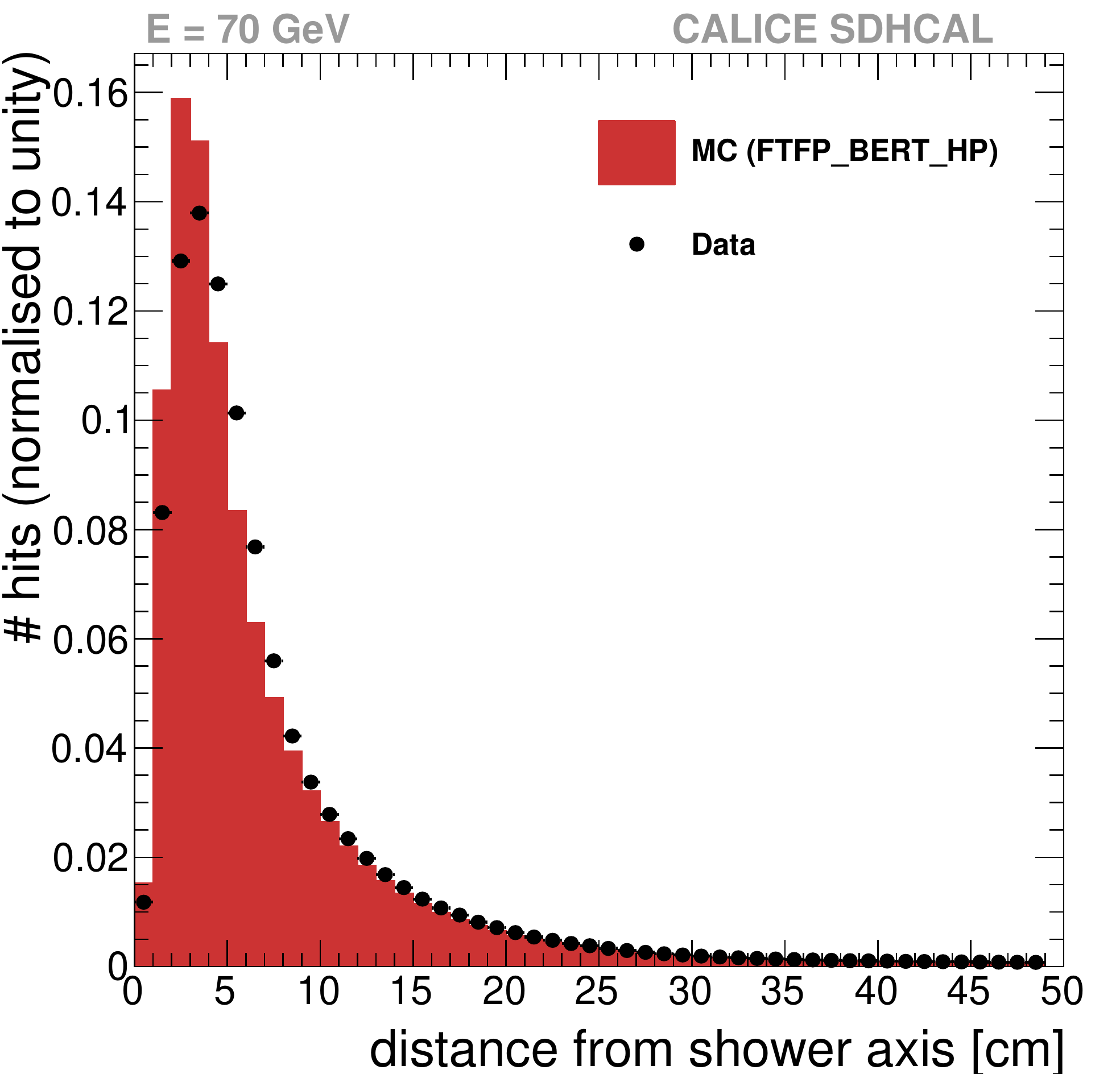}
  \caption{Radial shower profile for both data and simulation at 20 (left) and 70~GeV (right). }
  \label{fig.showerprofile}
\end{figure}
Figure~\ref{fig.showerprofile} presents comparisons between data and simulation of the radial shower profile for 20 and 70~GeV hadronic shower samples. The mean value $\langle R\rangle$ of the radial shower profile is defined as follows:
\begin{equation}
  \langle R\rangle=\frac{1}{N_{event}}\sum_{i=0}^{N_{event}}\sum_{r=0}^{R_{max}}r\frac{N_{r,i}}{N_{tot,i}}
\end{equation}
where $N_{event}$ is the number of events, $N_{r,i}$ is the number of hits in the ring of inner radius $r$ and $N_{tot,i}$ is the total number of hits for the event $i$. $R_{max}$ is the highest distance between the shower main thrust and a fired cell.
\begin{figure}[t]
  \center 
  \includegraphics[width=.45\textwidth]{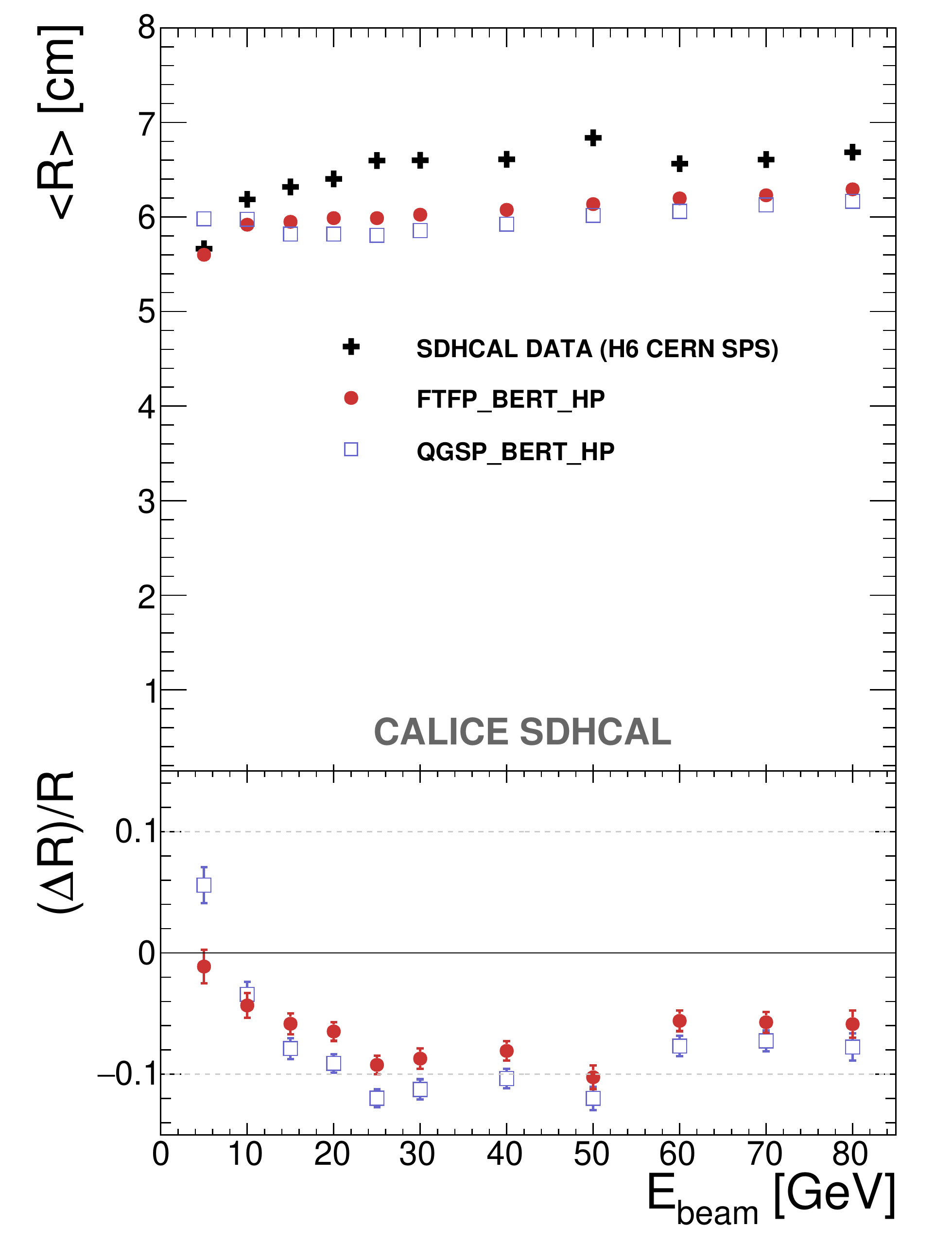}
  \caption{Mean value of the radial shower profile for pion runs as a function of the beam energy. Data are represented by black crosses, simulations are represented by red circles and open blue squares for FTFP\_BERT\_HP and QGSP\_BERT\_HP physics lists respectively. Relative deviations $\Delta R/ R$ are also presented (see text for details).}
  \label{fig.showerprofile_vs_energy}
\end{figure}
Figure~\ref{fig.showerprofile_vs_energy} shows $\langle R\rangle$ as function of the beam energy. Relative deviations (defined by $\frac{\langle \textmd{R}_{sim}\rangle - \langle \textmd{R}_{data} \rangle}{ \langle \textmd{R}_{data} \rangle}$) are also shown. For the two considered physics lists, the radial extent of hadronic showers is slightly underestimated. These results tend to confirm the previous conclusion on the radial shower profile~in~\cite{ahcal-geant}. 

%%%%%%%%%%%%%%%%%%%%%%%%%%%%%%%%%%%%%%%%%%%%%%%%%%%%%%%%%%%%%%%%%%%%%%%%%%%%%%%%%%%%%%%%%%%%%%%%%%

\section{Conclusion}
\label{sec:conclusion}
The SDHCAL simulation and the digitizer have been described. Simulation parameters have been extracted from data using response to incident muons and electrons. A good agreement between the data and the simulation on several variables such as efficiency, pad multiplicity extracted from muon samples and number of hits extracted from electromagnetic shower ones, suggests a reasonable description of the GRPC's response to charged particles. Although a detailed systematics study was not performed for this work, differences between the data and the simulation were observed on the number of hits with hadronic showers above 40~GeV in significant way. The number of reconstructed clusters, which is less dependent on the pad multiplicity, is also studied and it confirms the differences between data and simulation.
A topological variable, the radial shower profile, is also studied and found to be larger in data than in the simulation. This confirms independently of the digitizer the observed differences between data and simulation. It may explain the differences in number of hits mentioned above since larger radius means less saturation and thus more hits within SDHCAL.

\section{Acknowledgements}

We would like to thank the CERN-SPS staff for their availability and precious help during the two beam test periods.   We would like to acknowledge the important support provided by the  F.R.S.-FNRS, FWO (Belgium), CNRS and ANR (France), SEIDI and CPAN (Spain). This work was also supported by the Bundesministerium f\"{u}r Bildung und  Forschung (BMBF), Germany; by the Deutsche Forschungsgemeinschaft  (DFG), Germany; by the Helmholtz-Gemeinschaft (HGF), Germany; by  the Alexander von Humboldt Stiftung (AvH), Germany. 

%%%%%%%%%%%%%%%%%%%%%%%%%%%%%%%%%%%%%%%%%%%%%%%%%%%%%%%%%%%%%%%%%%%%%%%%%%%%%%%%%%%%%%%%%%%%%%%%%%

%\section*{References}

\end{document}

%% file: AuthorList.tex
\author{\centering 
\LARGE\bf The CALICE Collaboration
}

\author{\centering
Z.\,Deng,
Y.\,Li,
Y.\,Wang, 
Q.\,Yue, 
Z.\,Yang
\\ \it
Tsinghua University, Department of Engineering Physics.Beijing, 100084, P.R.
China
}

\author{\centering 
J.\,Apostolakis, 
G.\,Folger, 
C.\,Grefe$^1$,
V.\,Ivantchenko, 
A.\,Ribon,
V.\,Uzhinskiy
\\ \it 
CERN, 1211 Gen\`{e}ve 23, Switzerland}

\author{\centering
D.\,Boumediene, V.\,Fran\c cais
\\ \it
Universit\'e Clermont Auvergne, Universit\'e Blaise Pascal, CNRS/IN2P3, LPC, 4 Av. Blaise Pascal, TSA/CS 60026,
F-63178 Aubi\`ere, France
}

\author{\centering 
G.\,Cho, D-W.\,Kim, S.\,C.\,Lee, W.\,Park, S.\,Vallecorsa
\\ \it
Gangneung-Wonju National University, Gangneung 25457, South Korea
}
\author{\centering
S.\,Cauwenbergh, 
M.\,Tytgat,
A.\,Pingault,
N.\,Zaganidis
\\ \it
Ghent University, Department of Physics and Astronomy,
Proeftuinstraat 86, B-9000 Gent, Belgium
}

\author{\centering 
E.\,Brianne, 
A.\,Ebrahimi,
K.\,Gadow, 
P.\,G\"{o}ttlicher, 
C.\,G\"{u}nter, 
O.\,Hartbrich, 
B.\,Hermberg, 
A.\,Irles,
F.\,Krivan, 
K.\,Kr\"{u}ger, 
J.\,Kvasnicka, 
S.\,Lu, 
B.\,Lutz, 
V.\,Morgunov$^8$,
C.\,Neub\"{u}ser
A.\,Provenza,
M.\,Reinecke,  
F.\,Sefkow, 
S.\,Schuwalow,
H.L.\,Tran
\\ \it
DESY, Notkestrasse 85,
D-22603 Hamburg, Germany
}

\author{\centering  
E.\,Garutti, 
S.\,Laurien,
M.\,Matysek,
M.\,Ramilli, 
S.\,Schroeder
\\ \it
Univ. Hamburg,
Physics Department,
Institut f\"ur Experimentalphysik,
Luruper Chaussee 149,
22761 Hamburg, Germany
}

\author{\centering 
B.\,Bilki,
E.\,Norbeck$^8$,
D.\,Northacker,
Y.\,Onel
\\ \it
University of Iowa, Dept. of Physics and Astronomy,
203 Van Allen Hall, Iowa City, IA 52242-1479, USA
}

\author{\centering 
S.\,Chang, 
A.\,Khan, 
D.\,H.\,Kim, 
D.\,J.\,Kong, 
Y.\,D.\,Oh
\\ \it
Department of Physics, Kyungpook National University, Daegu, 702-701, 
Republic of Korea
}

\author{\centering 
K.\,Kawagoe, 
H.\,Hirai, 
Y.\,Sudo, 
T.\,Suehara, 
H.\,Sumida, 
T.\,Yoshioka
\\ \it
Department of Physics, Kyushu University, Fukuoka 819-0359, Japan
}

\author{\centering 
E.\,Cortina Gil,
S.\,Mannai
\\ \it
Center for Cosmology, Particle Physics and Cosmology (CP3)
Universit\'{e} catholique de Louvain, Chemin du cyclotron 2,
1320 Louvain-la-Neuve, Belgium
}

\author{\centering 
V.\,Buridon,
C.\,Combaret, 
L.\,Caponetto,
R.\,Et\'{e}, 
G.\,Garillot,
G.\,Grenier, 
R.\,Han$^2$
J.C.\,Ianigro,
R.\,Kieffer$^3$ ,
I.\,Laktineh,
N.\,Lumb, 
H.\,Mathez, 
L.\,Mirabito, 
A.\,Petrukhin,
A.\,Steen$^4$
\\ \it
Univ. Lyon, Universit\'{e} Lyon 1, 
CNRS/IN2P3, IPNL 4 rue E Fermi 69622,
Villeurbanne CEDEX, France
}

\author{\centering 
J.\,Berenguer~Antequera,
E.\,Calvo~Alamillo, 
M.-C.\,Fouz, 
J.\,Marin,
J.\,Puerta-Pelayo, 
A.\,Verdugo
\\ \it
CIEMAT, Centro de Investigaciones Energeticas, Medioambientales y Tecnologicas, Madrid, Spain 
}

\author{\centering 
M.\,Chadeeva$^9$,
M.\, Danilov$^{10}$,
\\ \it
National Research Nuclear University 
MEPhI (Moscow Engineering Physics Institute)
31, Kashirskoye shosse,
115409 Moscow, Russia
}

\author{\centering 
M.\,Gabriel, 
P.\,Goecke,
C.\,Kiesling,
N.\,van\,der\,Kolk, 
F.\,Simon, 
M.\,Szalay, 
\\ \it
Max Planck Inst. f\"ur Physik,
F\"ohringer Ring 6,
D-80805 Munich, Germany
}

\author{\centering 
S.\,Bilokin,
J.\,Bonis, 
P.\,Cornebise, 
F.\,Richard, 
R.\,P\"oschl, 
J.\,Rou\"en\'e, 
A.\,Thiebault,
D.\,Zerwas
\\ \it
Laboratoire de L'acc\'elerateur Lin\'eaire,
Centre d'Orsay, Universit\'e de Paris-Sud XI,
BP 34, B\^atiment 200,
F-91898 Orsay CEDEX, France
}

\author{\centering 
M.\,Anduze,
V.\,Balagura,
K.\,Belkadhi,
V.\,Boudry, 
J-C.\,Brient, 
R.\,Cornat,
M.\,Frotin,
F.\,Gastaldi, 
Y.\,Haddad$^5$, 
F.\,Magniette,
M.\,Ruan$^6$,
M.\,Rubio-Roy,
K.\,Shpak,
H.\,Videau,
D.\,Yu$^{11}$
\\ \it
 Laboratoire Leprince-Ringuet (LLR)  -- \'{E}cole polytechnique, CNRS/IN2P3, Palaiseau, F-91128 France
}

\author{\centering   
S.\,Callier,
S.\, Conforti di Lorenzo, 
F.\,Dulucq, 
G.\,Martin-Chassard, 
Ch.\,de la Taille, 
L.\,Raux, 
N.\,Seguin-Moreau
\\  \it
 Laboratoire OMEGA  -- \'{E}cole Polytechnique,
 CNRS/IN2P3,
 Palaiseau, F-91128 France
}

\author{\centering              
K.\,Kotera, 
H.\,Ono$^7$,%\footnote{}, 
T.\,Takeshita
\\ \it
Shinshu Univ.\,,
Dept. of Physics,
3-1-1 Asaki,
Matsumoto-shi, Nagano 390-861,
Japan
}

\author{\centering 
F.\,Corriveau
\\ \it
Department of Physics, McGill University,
Ernest Rutherford Physics Bldg.,
3600 University Ave.,
Montr\'{e}al, Quebec,
CANADA H3A 2T8
}

\author{
\it
$^\star$ Corresponding author\newline
E-mail: \email{arnaud.steen@cern.ch}
}

\author{\\
  \llap{$^1$} Now at University of Bonn, Bonn.\\
  \llap{$^2$} Now at CAST, Beijing.\\
  \llap{$^3$} Now at CERN, Geneva.\\
  \llap{$^4$} Now at National Taiwan University, Taipei.\\
  \llap{$^5$} Now at Imperial College, London.\\
  \llap{$^6$} Now at IHEP, Beijing.\\
  \llap{$^7$} Now at Nippon Dental University, Japan.\\
  \llap{$^8$} Deceased.\\
  \llap{$^9$} Also at P.\,N.\, Lebedev Physical Institute.\\
  \llap{$^{10}$} Also at P.\,N.\, Lebedev Physical Institute and in MPTI.\\
  \llap{$^{11}$} Also at at IHEP, Beijing.\\
}